\documentclass[preprintnumbers,amsmath,amssymb,floatfix,10pt,prd,onecolumn,layout,
superscriptaddress,nofootinbib,conference]{revtex4}
\usepackage{latexsym}
\usepackage{epsfig}
\usepackage{epstopdf}
\usepackage{graphicx}
\usepackage{amssymb}
\usepackage{amsmath}
\usepackage{dcolumn}
\usepackage{bm}
\usepackage{color}
\usepackage{comment}
\usepackage{float}
\begin{document}
\title{\bf The Optical Appearance of Charged Four-Dimensional
Gauss-Bonnet Black Hole with Strings Cloud and Non-Commutative
Geometry Surrounded by Various Accretions Profiles}
\author{Xiao-Xiong Zeng}
\altaffiliation{xxzengphysics@163.com}\affiliation{State Key
Laboratory of Mountain Bridge and Tunnel Engineering, Chongqing
Jiaotong University, Chongqing 400074, China}\affiliation{Department
of Mechanics, Chongqing Jiaotong University, Chongqing 400074,
China}
\author{M. Israr Aslam}
\altaffiliation{mrisraraslam@gmail.com}
\author{Rabia Saleem}
\altaffiliation{rabiasaleem@cuilahore.edu.pk}\affiliation{Department
of Mathematics, COMSATS  University Islamabad, Lahore Campus,
Pakistan.}

\begin{abstract}
Thanks for the releasing image of supermassive black holes (BHs) by
the event horizon telescope (EHT) at the heart of the $M87$ galaxy.
After the discovery of this mysterious object, scientists paid
attention to exploring the BH shadow features under different
gravitational backgrounds. In this scenario, we study the light
rings and observational properties of BH shadow surrounded by
different accretion flow models and then investigate the effect of
model parameters on the observational display and space-time
structure of BHs in the framework of our considering system. Under
the incompatible configuration of the emission profiles, the images
of BHs comprise that the observed luminosity is mainly determined by
direct emission, while the lensing ring will provide a small
contribution of the total observed flux and the photon ring makes a
negligible contribution due to its exponential narrowness. More
importantly, the observed regions and specific intensities of all
emission profiles are changed correspondingly under variations of
parameters. For optically thin accreting matters, we analyze the
profile and specific intensity of the shadows with static and
infalling accretions models, respectively. We find that with an
infalling motion the interior region of the shadows will be darker
than in a static case, due to the Doppler effect of the infalling
movement. Finally, it is concluded that these findings support that
the change of BH state parameters will change the way of space-time
geometry, thus affecting the BH shadows dynamics.
\end{abstract}
\date{\today}
\maketitle

\section{Introduction}

A black hole is a mysterious region of space-time, where the
gravitational force is extremely intense that nothing, no particles,
or even electromagnetic radiation such as light, can escape from it.
A BH can be formed at the final stage of a massive star. When such a
star has completely exhausted the nuclear fuel in its core at the
end of its life cycle, the core becomes unbalanced and gravitational
collapse occurs inside the core, and the outer layers of the star
are blown away. After this, the massless weight of constituent
matter falls in the dying sphere to a point of zero volume and
infinite density, so-called singularity. The simplest model for BH
formation involves a collapsing thin spherical shell of massless
matter, i.e., a shell of photons, gravitons, or massless neutrinos
with very small radial extension and total energy. In general
relativity (GR), BH is one of the most fascinating predictions, and
researchers have been trying to resolve this mysterious puzzle. The
Laser-Interferometer Gravitational Wave-Observatory (LIGO) detects
the emission of gravitational waves from BHs merger \cite{1}.

In 2019, the EHT captured the first evidence for the existence of
ultra-high angular resolution image of accretion flows around a
super-massive object at the center of $M87$
($M=6.5\pm0.7\times10^{9}M_{\odot}$) galaxy \cite{2,3,4,5,6,7}. The
canonical interpretation of the $M87$ galaxy shows that the
gravitational field of a BH contains a photon sphere when generally
illuminated by an thin accretion matter. The interior of dark region
is bounded by a bright band, known as the BH shadow and photon
sphere, respectively. The existence of a strong gravitational
interaction at the core of BH regimens makes the gravitational
deflection of light, which provide a comprehensive way to analyze
the BH shadow. Therefore, the analysis of BH shadow and
gravitational lensed may provide the feasible way to evaluate the
strength of the gravitational field in space-time geometry and
support the prediction of Einstein's GR \cite{8,9,10,11}.

The intensity of emitted light depends upon the location of the
distant observer, leading to a dark inner region and intense photon
ring. Generally, the shadow of different BHs has been discussed in
the literature and tries to find a complex pattern of several light
ray trajectories around a dark thin disk. The study of light
deflection by a compact star or BH was initially studied in
\cite{9}. Later, the extension of BH thin accretion disk and
critical curve observed the image of BH in \cite{11}. Theoretically,
the critical curve is the light trajectory traced backward from the
distant viewer, which would have asymptotically approach towards a
bound photon ring. For the Schwarzschild BH, the value of bound
photon ring is $r=3M$, (where $M$ is the mass of BH) and the radius
of the critical impact parameter curve is $b=3\sqrt{3}M$.

However, the region of impact parameter depends on the geometrical
interpretation and various physical properties of the illuminating
accretion flow of BHs \cite{12}. In addition, the width and the
intensity of lensing rings is changed with the variation of emission
region and size of the shadow dependent on the emission model, the
accretion has a minor effect on the dark central region \cite{13}.
The influence of quintessence dark energy dynamics on the shadow of
BH and thin accretions disk surrounded by various trajectories of
light are widely discussed in \cite{14}. Further, the authors in
\cite{14} concluded that the existence of a cosmological horizon
plays an significant part in the BH shadows and the location of
photon rings for both stationary and infalling accretions are lie in
the same orbit. Shadows and photon rings with the static and
spherical infalling accretions are analyzed in the framework of
four-dimensional Gauss-Bonnet (GB) BH, and the influence of the GB
coupling constant on the BH shadow and dynamics of photon spheres
have been calculated in \cite{15}.

In the background of Einstein GB-Maxwell gravity, Ma et al.
\cite{16} studied the spherically symmetric charged BH shadow and
its photons sphere with the sequence of inequalities of the BH
horizon and its mass. Gao et al. \cite{17} discussed gravitational
lensing of a hairy BH in Einstein-Scalar-GB gravity and compared the
consistency of BH shadow with EHT data. Guo et al. \cite{18}
considered the perfect fluid within Rastall gravity and investigated
the shadow and photon sphere of charged BH with infalling accretion
and obtained brighter photon sphere luminosity than the static
spherical accretion. The luminosity of shadows and light rings of
the Hayward charge BH are affected by the accretion flow property is
a result obtained in \cite{19}. The analysis of BH shadow in GR as
well as in extended gravitational theories provides a new way to
discover the properties of BHs \cite{20,21,22,23,24,25,26,27}.

The idea of the non-commutative (NC) geometry is extensively used to
evaluate the BH solutions, wormhole geometry, and cosmological
constraints. Recently, NC geometry has obtained a significant
attraction as it provides an comprehensive way to analysis the
effects of quantum gravity in space-time structure \cite{f1,f2}. For
instance, the influence of NC operators on BH is a interesting
subject, and a number of methods are being proposed to explore the
NC geometry \cite{f3,f4,f5}. This geometry can be implemented to GR
by modifying the matter source, considering the minimal length
instead of the Dirac function, replaced by Gaussian distribution or
Lorentzian distribution \cite{29}. Regarding the growing interest of
researchers in further analysis of BH shadows, we have proposed a
new approach to discussing the observational characteristics of BH
space-time. In the present manuscript, we investigate the shadow of
four-dimensional GB charged BH with spherical accretions under the
influence of a cloud of strings and NC geometry \cite{27,28,29}.
Particularly, we study the qualitative features of BH shadows under
the influence of charge, the cloud of strings, and NC geometry
parameters. In Sec. \textbf{II}, we provide the basic formulation of
our considering system and studied the effective potential
corresponding to the light trajectories. Section \textbf{III} is
dedicated to the optical appearance of photon rings with thin disk
accretion flow models and to studying the shadow with specifically
observed intensities. The analysis of shadows and photon spheres
rings with a static accretion matter is dealt with in Sec.
\textbf{IV}. In next section, we discuss the dynamics of the BH
accretion with infalling matters. The last section is devoted to the
conclusion and discussion of the current analysis.

\section{Light Deflection in the Charged GB BH with Cloud of Strings
and NC Geometry}

The Einstein-Hilbert action of GB gravity is formulated as \cite{15}
\begin{equation}\label{1}
\mathcal{I}=\frac{1}{16 \pi G}\int
d^4{x\sqrt{-g}}[R+\alpha(R_{\gamma\delta\zeta\xi}R^{\gamma\delta\zeta\xi}-4R_{\gamma\delta}R^{\gamma\delta}+R^{2})],
\end{equation}
where $R$ is the curvature scalar, by re-scaling the GB coupling constant $\alpha$, i.e., $\alpha$/$D-4$ and taking the limit
$D\rightarrow 4$ in the GB term, one can obtain the solution of four-dimensional GB BH \cite{30}.
We consider the static and spherically symmetric metric for four-dimensional GB BH \cite{28}
\begin{equation}\label{2}
ds^{2}=-f(r)dt^{2}+\frac{dr^{2}}{f(r)}+r^{2}d\theta^{2}+r^{2}\sin^{2}\theta
d\varphi^{2},
\end{equation}
with
\begin{equation}\label{3}
f(r)=1+\frac{r^{2}}{2\alpha}\bigg(1-\sqrt{1+4\alpha\big(\frac{2M}{r^{3}}-\frac{Q}{r^{4}}+\frac{a}{r^{2}}\big)}\bigg).
\end{equation}
The solution (\ref{3}) can be characterized by the GB coupling constant $\alpha$, mass $M$, Charge $Q$ and cloud of string
parameter $a$, which is consider to be positive.

The point-like structure with smeared objects, the mass density of a
static and spherically symmetric gravitational source is given by
Lorentzian distribution as follows \cite{29}
\begin{equation}\label{4}
\rho_{\phi}=\frac{\sqrt{\phi}M}{\pi^{3/2}(\pi\phi+r^{2})^{2}},
\end{equation}
where $\phi$ is the strength of NC parameter in
Lorentzian distribution. The smeared mass of matter distribution can be obtain as \cite{27}
\begin{eqnarray}\nonumber
\mathcal{M_{\phi}}&=&\int^{r}_{0}4\pi
r^{2}\rho_{\phi}(r)dr=\frac{2M}{\pi}\bigg[\arctan\big(\frac{r}{\sqrt{\pi
\phi}}\big)-\frac{r\sqrt{\pi\phi}}{\pi\phi+r^{2}}\bigg],\\\label{5}&=&M-\frac{4\sqrt{\phi}M}{\sqrt{\pi}r}+\mathcal{O}(\phi^{3/2}).
\end{eqnarray}
In this way, the equation (\ref{3}) can be rewritten as
\begin{equation}\label{6}
f(r)=1+\frac{r^{2}}{2\alpha}\bigg(1-\sqrt{1+4\alpha\big(\frac{2\mathcal{M}_{\phi}}{r^{3}}-\frac{Q}{r^{4}}+
\frac{a}{r^{2}}\big)}\bigg),
\end{equation}
which modify the NC BH geometry as
\begin{equation}\label{7}
f(r)=1+\frac{r^{2}}{2\alpha}\bigg(1-\sqrt{1+4\alpha\big(\frac{2M}{r^{3}}-\frac{8\sqrt{\phi}M}{\sqrt{\pi}r^{4}}-\frac{Q}{r^{4}}+
\frac{a}{r^{2}}\big)}\bigg).
\end{equation}

In order to found the location of horizons, one can solve the
equation $f(r)=0$, and following two analytic solutions are obtained
\begin{eqnarray}\label{8}
r_{h}&=&\frac{M\pi^{1/4}+\sqrt{\sqrt{\pi}(M^{2}+(a-1)((Q+\alpha)+8M\sqrt{\phi}))}}{(1-a)\pi^{1/4}},\\\label{9}
r_{c}&=&\frac{M\pi^{1/4}-\sqrt{\sqrt{\pi}(M^{2}+(a-1)((Q+\alpha)+8M\sqrt{\phi}))}}{(1-a)\pi^{1/4}},
\end{eqnarray}
where $r_{h}$ and $r_{c}$ correspond to the event horizon and
cosmological horizon of the BH, respectively. For the existence of a
horizon, the GB coupling parameter $\alpha$ should fall in the
allowed range $-8\leq\alpha/M\leq1$. For $\alpha>0$, we have two
horizons, while for $\alpha<0$, only one horizon exists. Although
$\alpha>0$ depicts the properties of the inverse string tension, the
solution (\ref{3}) also allows the case $\alpha<0$. So, we expect
some interesting aspects of GB gravity as it was argued in
\cite{30}.

To analysis the accretion flow of photons spheres around the BH, we
need to analyze the behavior of light deflection in geometrical
optics near the BH shadow. The geodesic motion can be encapsulated
with the help of the Euler-Lagrange equation, can be written in the
following form
\begin{equation}\label{10}
\frac{d}{d\eta}\bigg(\frac{\partial\mathcal{L}}{\partial
\dot{x}^{\alpha}}\bigg)=\frac{\partial\mathcal{L}}{\partial
x^{\alpha}},
\end{equation}
in which $\eta$ is the affine parameter, $\dot{x}^{\alpha}$ is the
four-velocity of BH light rays, ``$.$'' represents the derivative
with respect to $\eta$. The Lagrangian $(\mathcal{L})$ of photons can
be written as:
\begin{eqnarray}\label{11}
\mathcal{L}=\frac{1}{2}g_{\alpha\beta}\dot{x}^{\alpha}\dot{x}^{\beta}=
\frac{1}{2}\bigg(-f(r)\dot{t}^{2}+\frac{\dot{r}^{2}}{f(r)}+r^{2}\dot{\theta}^{2}+r^{2}\sin^{2}\theta
\dot{\varphi}^{2}\bigg).
\end{eqnarray}
The spherical symmetry allows us to choose the motion of photons on
the equatorial plane with $\theta=\pi/2$ and $\dot{\theta}=0$ without
loss of generality \cite{10,31}. From (\ref{2}), the metric
coefficients are independent of time $t$, and the azimuthal angle
$\varphi$. So, two conserved quantities can be evaluated
such as energy $E=\partial\mathcal{L}/\partial \dot{t}=f(r)\dot{t}$,
and angular momentum of photon
$\mathcal{J}=\partial\mathcal{L}/\partial
\dot{\varphi}=r^{2}\dot{\varphi}$. Using Eqs. (\ref{7}),
(\ref{10}) and (\ref{11}), we obtain
\begin{equation}\label{12}
\dot{t}=\frac{1}{b\bigg(1+\frac{r^{2}}{2\alpha}\bigg(1-\sqrt{1+4\alpha\big(\frac{2M}{r^{3}}-
\frac{8\sqrt{\phi}M}{\sqrt{\pi}r^{4}}-\frac{Q}{r^{4}}+
\frac{a}{r^{2}}\big)}\bigg)\bigg)},
\end{equation}
\begin{equation}\label{13}
\dot{\varphi}=\pm\frac{1}{r^{2}},
\end{equation}
\begin{equation}\label{14}
\dot{r^{2}}+\frac{1}{r^{2}}\bigg(1+\frac{r^{2}}{2\alpha}\bigg(1-\sqrt{1+4\alpha\big(\frac{2M}{r^{3}}-
\frac{8\sqrt{\phi}M}{\sqrt{\pi}r^{4}}-\frac{Q}{r^{4}}+
\frac{a}{r^{2}}\big)}\bigg)\bigg)=\frac{1}{b^{2}},
\end{equation}
where $``+"$ and $``-"$ in Eq. (\ref{13}) corresponds to the
anti-clockwise/clockwise motion of photon, respectively. Further, we
define the affine parameter $\eta$, by $\frac{\eta}{\mathcal{J}}$
and the impact parameter $b_{c}=\frac{\mathcal{J}}{E}$, which is
used to evaluate the vertical distance between the two lines, such
as geodesic and parallel lines having the same origin. Equation
(\ref{14}) is used to evaluate the geodesic equation in the form of
the effective potential, one can write as
\begin{equation}\label{15}
\dot{r^{2}}+V_{e}(r)=\frac{1}{b^{2}},
\end{equation}
where
\begin{equation}\label{16}
V_{e}(r)=\frac{1}{r^{2}}\big(f(r)\big).
\end{equation}
In the equatorial plane, the null geodesic exists in space-time
region and the light rays projected on the equatorial plane,
yielding a circular orbit. The position of the maximum effective
potential correspond to the frequency of threshold stability for the
null geodesic circular geometry around $b_{c}$ having a critical
curve. The photon will move around the BH in an unbalanced circular
orbit, at this time, the surface of the photon sphere corresponds to
the surface of the circular orbit. The motion of the light rays on
the photon sphere should satisfy the conditions $\dot{r}=0$ and
$\ddot{r}=0$. Further, the motion of photon sphere orbit can be
translated as
\begin{eqnarray}\label{17}
V_{e}(r)=\frac{1}{b^{2}}, \quad  V'_{e}(r)=0,
\end{eqnarray}
where prime represent the derivative with respect to $r$. Based on
this equation, the shadow radius of photon sphere $r_{p}$, impact
parameter of the critical curve $b_{c}$ and radii of event horizon
$r_{h}$ are obtained for different values of model parameters and
listed in Table. \textbf{1}. The quantities $r_{p}$ and $b_{c}$ are
satisfying the following equations:
\begin{eqnarray}\label{18}
r^{2}_{p}=b^{2}_{c}f(r), \quad 2b^{2}_{c}f(r)^{2}-r^{3}_{p}f'(r)=0.
\end{eqnarray}
Due to the correlation of parameters, the geometry of space-time varies
with the variation of parameters, which means that the trajectories
of photons will be different corresponding to different numerical values. For
instant, we choose fixed values of charge $Q$ and NC
parameter $\phi$ and evaluate the event horizon $r_{h}$, critical
curve $b_{c}$ and radius of photon sphere $r_{p}$, for different
values of coupling constant $\alpha$ and cloud of strings parameter
$a$. Similarly, we can also evaluate the numerical values of
$r_{h}$, $b_{c}$ and $r_{p}$ corresponding to different values $Q$
and $\phi$ by fixing other parameters.

One can see from Table. \textbf{1}, the numerical values of $r_{h}$
, $b_{c}$ and $r_{p}$ decreases with an increase in the values of
$\alpha$. On the other hand, these values increase directly with
increasing values of $a$. Hence, it implies that the BH photon ring
shrinks inward as one increases the values of $\alpha$ and expands
outward concerning $a$. Taking some fixed values of the parameters,
we depict the effective potential $V_{e}(r)$, in Fig. \textbf{1}
(left panel). Let us consider Fig. \textbf{1} (left panel), there is
no effective potential at the event horizon. The trajectory of
effective potential increases maximum in the position of the photon
sphere and then vanishes gradually with the increment of radius $r$.
\begin{table}[H]
\label{table:6} \centering \small\small \begin{tabular}{|c| c| c| c|
c| c| c| c| c| c| c| c| c| c|}
 \hline
  \textbf{$\alpha$} & $-7.55$& $-5.55$& $-2.55$ &$0.22$ & $0.44$ & $0.55$ \\ [0.5ex]
 \hline
 $r_{h}$ & $ 4.11292 $ & $3.71661$ & $2.96995 $ & $ 1.72553 $ & $ 1.47589 $ & $1.21524$ \\\hline
 $b_{c}$ & $ 7.70838 $ & $7.31300$& $ 6.55952 $ & $ 5.37765 $ & $ 5.20921 $ & $5.11003$ \\\hline
 $r_{p}$ & $ 5.01684 $ & $4.64989$& $ 3.94120 $ & $ 2.75860 $ & $ 2.56800$ &  $2.44825$
 \\\hline\hline
\textbf{$a$}& $0.1$& $0.2$& $0.3$&$0.4$ & $0.5$ & $0.6$
\\ [0.5ex]
 \hline
 $r_{h}$ & $ 1.61637 $ & $1.92883$ & $ 2.13273 $ & $ 2.81072 $ & $ 3.49576 $ & $4.51162$ \\\hline
 $b_{c}$ & $ 5.29741 $ & $6.42990$ & $ 7.98517 $ & $ 10.22250 $ & $ 13.64610 $ & $19.36120$ \\\hline
 $r_{p}$ & $ 2.66938 $ & $3.10145$ & $ 3.64973 $ & $ 4.37446 $ & $ 5.38328 $ & $6.89082$ \\ [1ex]
 \hline
\end{tabular}
\caption{The numerical values of involved physical quantities for
various values of $\alpha$ and $a$ for fixed values of $Q=0.1$ and $\phi=0.01$
with $M=1$.}
\end{table}
\begin{figure}[thpb]
\centering
\includegraphics[width=5.9cm,height=4.6cm]{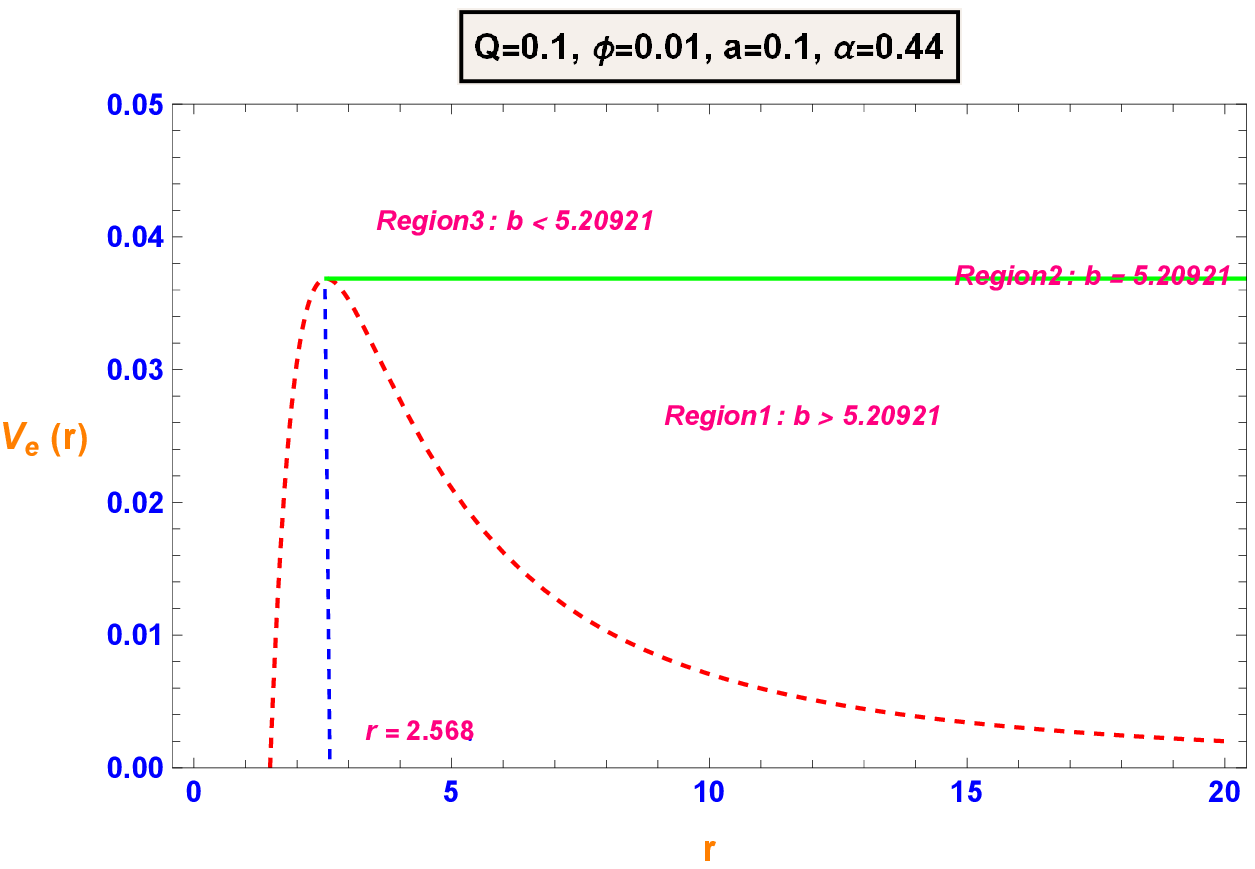}
\includegraphics[width=5.9cm,height=4.6cm]{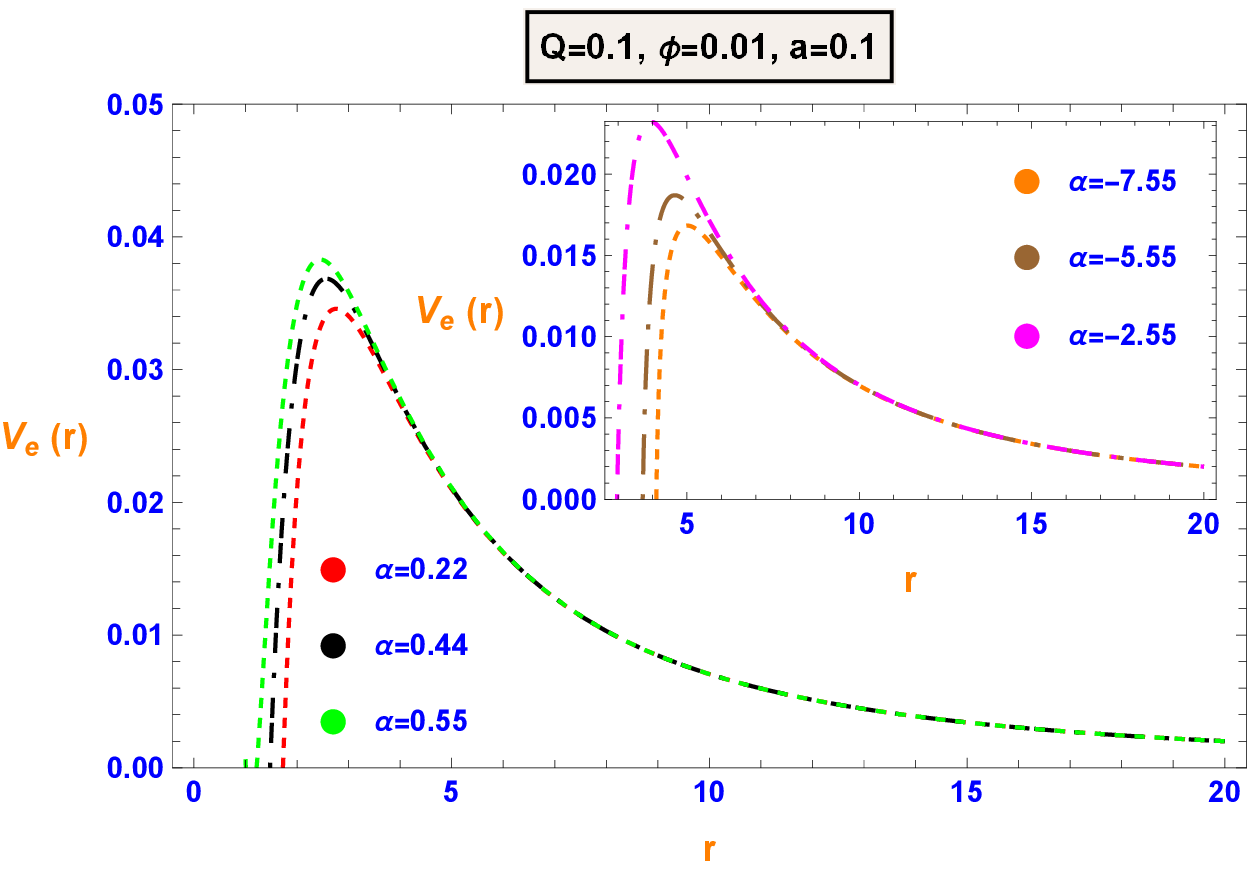}
\includegraphics[width=5.9cm,height=4.6cm]{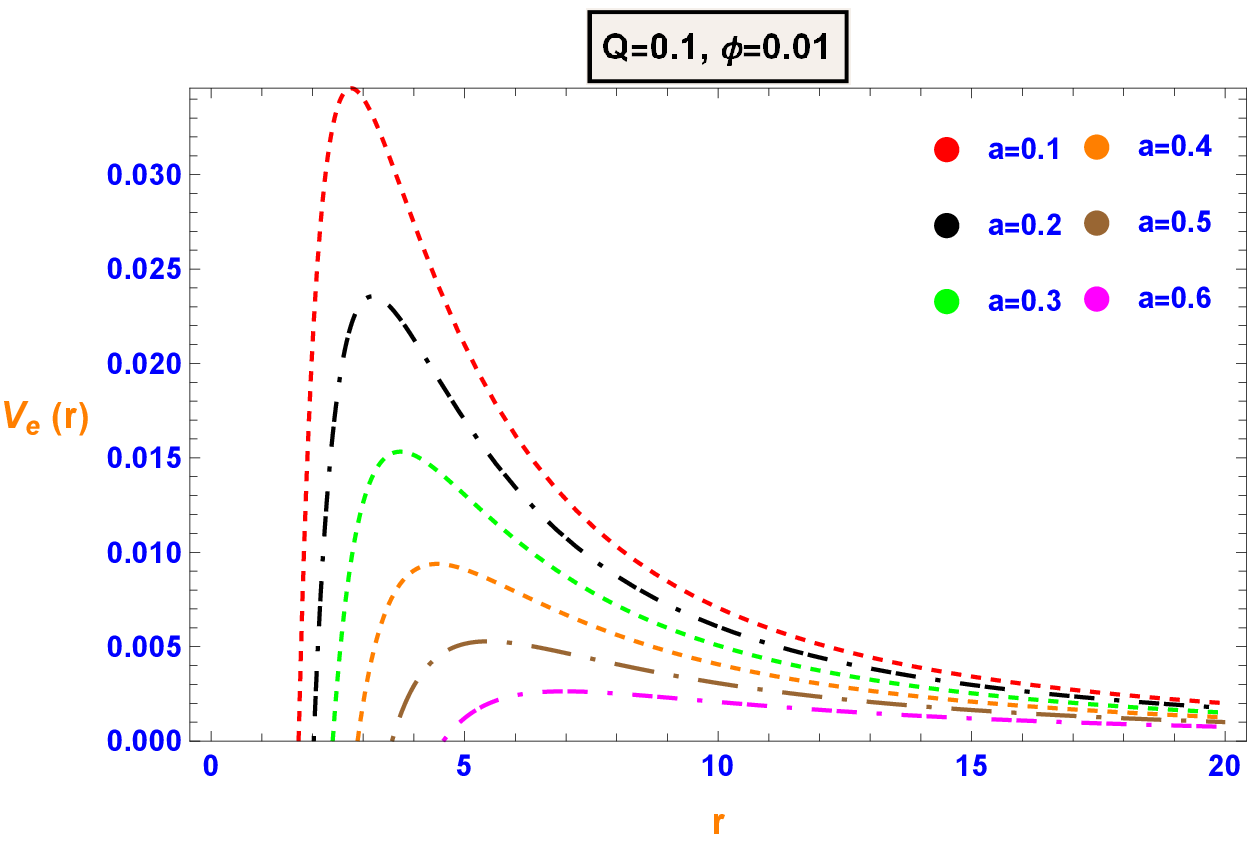}
\caption{The Behaviour of $V_{e}(r)$ versus $r$ for some specific
values of model parameters (left panel), for different $\alpha$
(middle panel) and for different $a$ (right panel) with $M=1$. In
left panel, dashed blue line indicates the radius of photon sphere
$r_{p}$ while green line corresponds to $V_{e}(r)=1/b_{c}^{2}$.}
\end{figure}
In region $1$ ($b>b_{c}$), the light rays hit the potential barrier
which is generated due to $V_{e}(r)$, region $2$ ($b=b_{c}$) defines
the behavior of impact parameter which is nearest to the photon
sphere radius and rotate around the BH, and region $3$ ($b<b_{c}$),
light fall into BH because there is no encounter the potential
barrier. The changing trend of the effective potential for different
values of $\alpha$ and $a$ are presented in Fig. \textbf{1} (middle
and right panels). To depict the geodesics of photons, using
Eqs. (\ref{13}) and (\ref{15}) with the setting of $u_{0}=1/r$ as
\begin{eqnarray}\label{19}
\Psi(u_{0})=\frac{du_{0}}{d\varphi}=\bigg(\frac{1}{b^{2}}-u_{0}^{2}\bigg(1+\big
(1-\big(1+4\alpha(au_{0}^{2}+2Mu_{0}^{3}-Qu_{0}^{4}-(8Mu_{0}^{4}\sqrt{\phi})/\sqrt{\pi})
\big)^{\frac{1}{2}}\big)/2u_{0}^{2}\alpha\bigg)\bigg)^{\frac{1}{2}}.
\end{eqnarray}
The optical appearance of the BH shadow depends on Eq. (\ref{19}),
and we plot the ingoing or outgoing trajectories through the
ray-tracing procedure as presented in Figs. \textbf{2} and
\textbf{3}. The light deflection for the region $b<b_{c}$ (orange
lines), the light rays fall into BH. For case, $b=b_{c}$ (green
lines), locate the position of the photon sphere and revolve around
the BH whereas for $b>b_{c}$ (blue lines), the light ray deflected
and move towards the BH from an infinite location to one closest
point and moves away from BH to infinity.

In addition, regions $1$, $2$ and $3$ are presented in Fig.
\textbf{1} (left panel) correspond to the blue, green, and orange
lines in Figs. \textbf{2} and \textbf{3} in general sense. We plot
Figs. \textbf{2} and \textbf{3} for some specific choices of model
parameters and find the shadow image of BH in space-time which is
different for each set of these values.
\begin{figure}[thpb]\centering
\includegraphics[width=5.9cm,height=5.6cm]{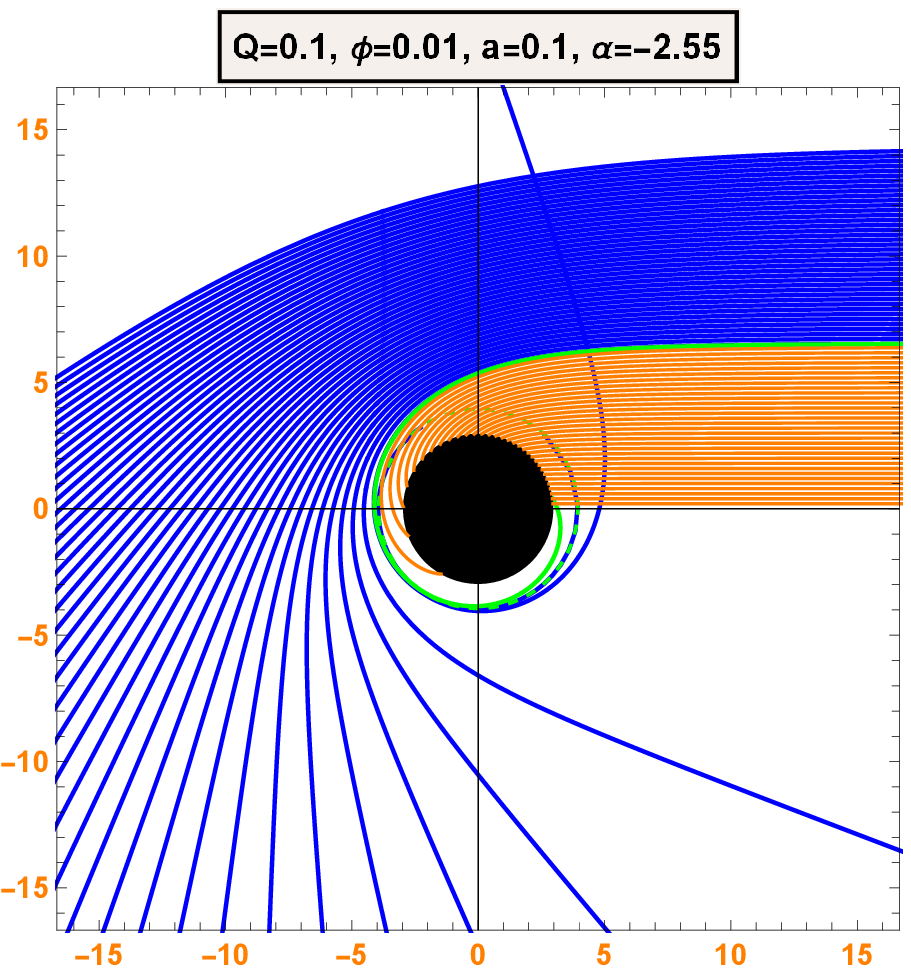}
\includegraphics[width=5.9cm,height=5.6cm]{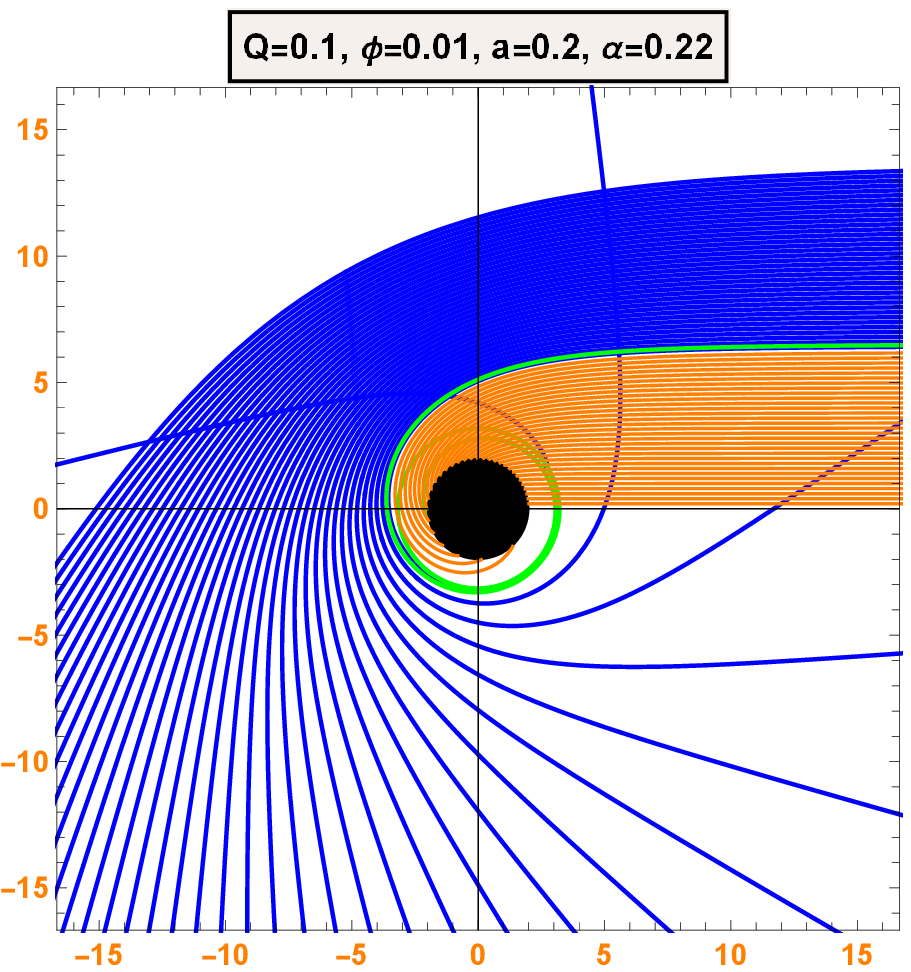}
\includegraphics[width=5.9cm,height=5.6cm]{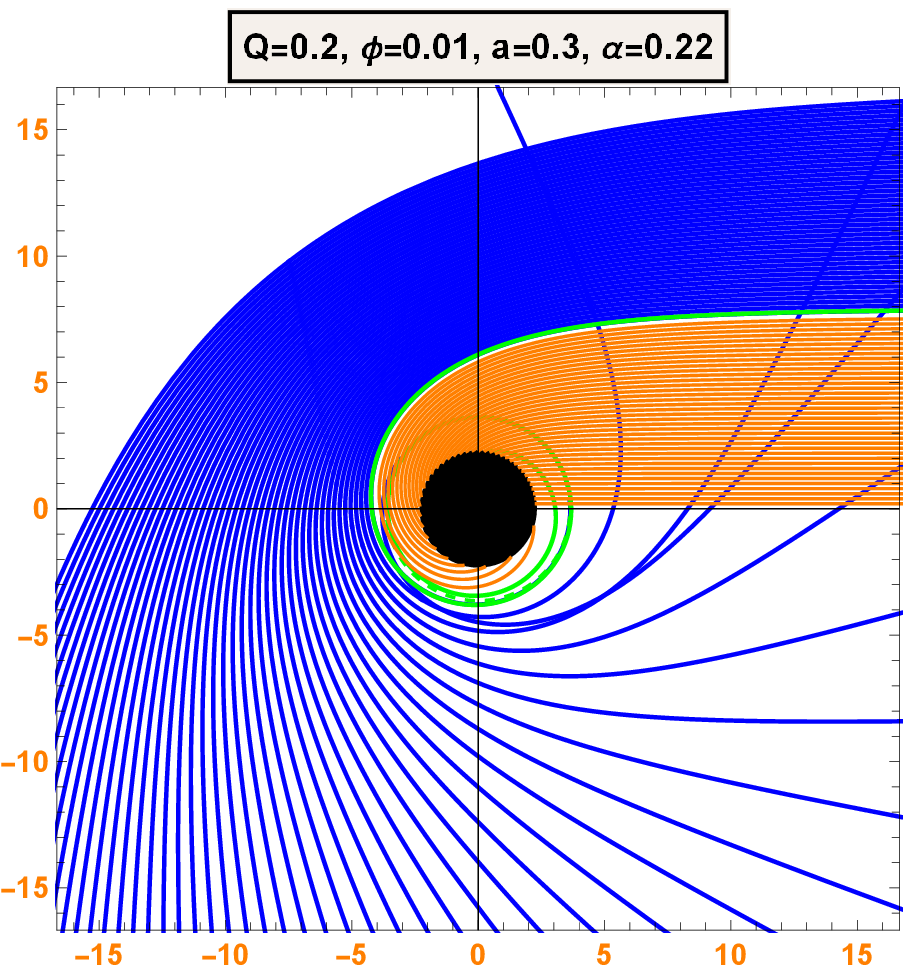}
\caption{The trajectories of photon rings for different parameters
in polar coordinates $(r,\varphi)$. The considering BH is shown as
solid black disks, and the blue, green and orange lines correspond
to $b>b_{c}$, $b=b_{c}$ and $b<b_{c}$ regions, respectively. The
dashed green line represents the radius of photon sphere and the
mass of BH as $M=1$.}
\end{figure}
\begin{figure}[thpb]\centering
\includegraphics[width=5.9cm,height=5.6cm]{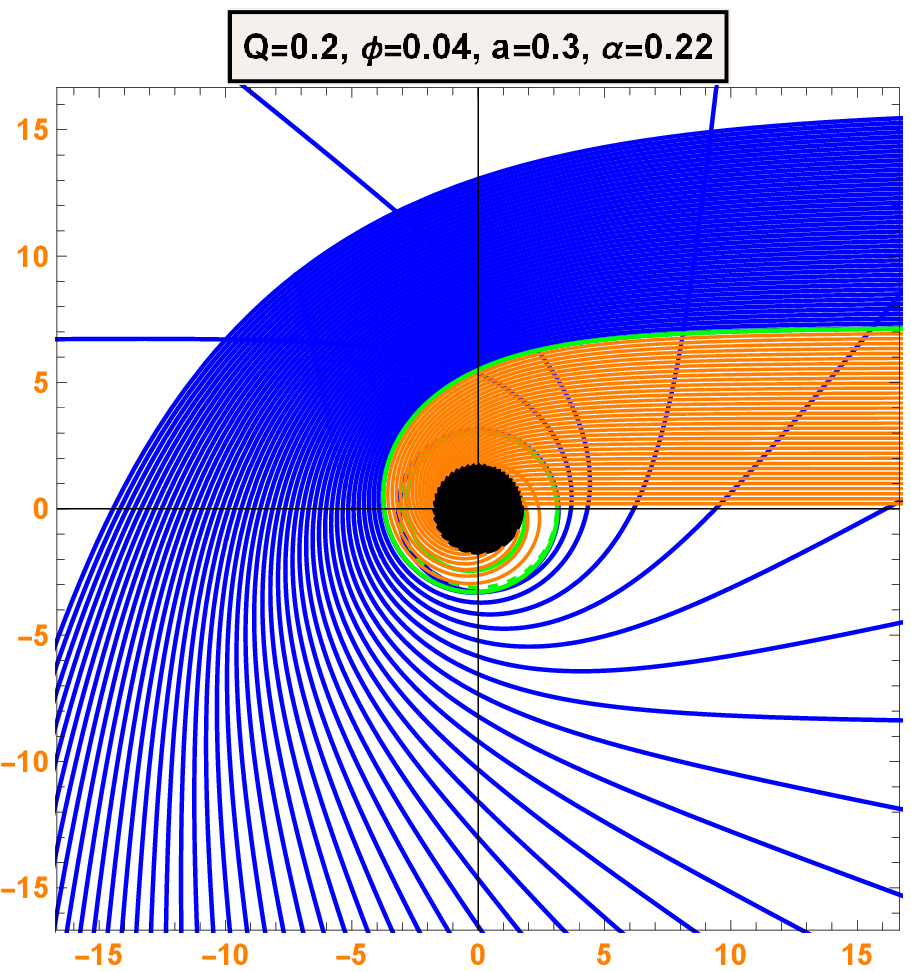}
\includegraphics[width=5.9cm,height=5.6cm]{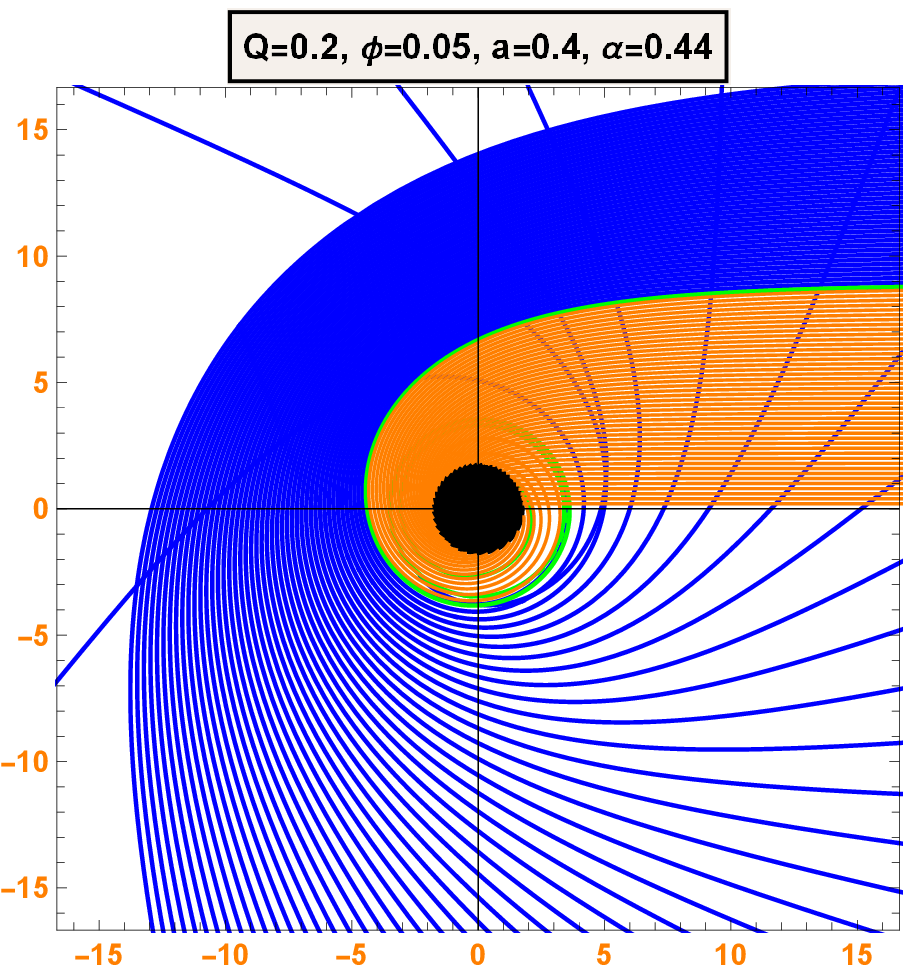}
\includegraphics[width=5.9cm,height=5.6cm]{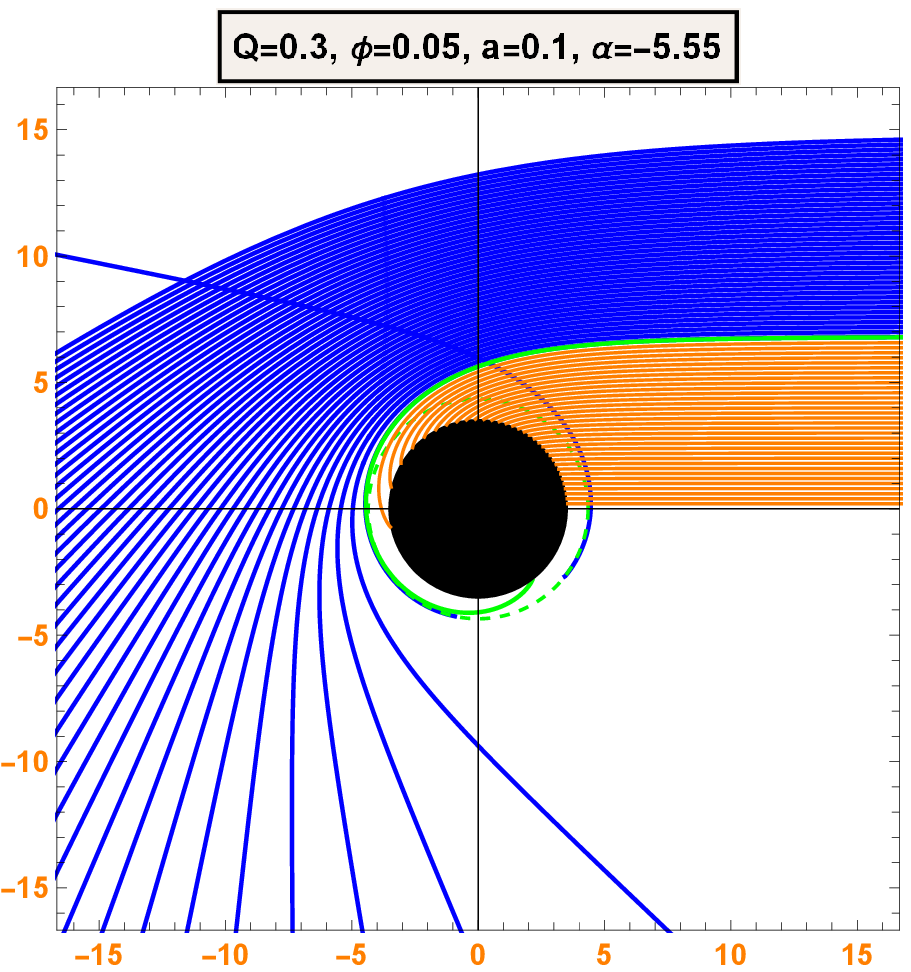}
\caption{The trajectories of photon rings for different parameters
in polar coordinates $(r,\varphi)$. The considering BH is shown as
solid black disks, and the blue, green and orange lines correspond
to $b>b_{c}$, $b=b_{c}$ and $b<b_{c}$ regions, respectively. The
dashed green line represents the radius of photon sphere and the
mass of BH as $M=1$}
\end{figure}

\section{Shadows and Photon Rings with Thin-Accretion Flow Models}

Now, we are going to discuss the optical appearance of thin disk
accretion around the BH in NC geometry with the cloud of strings and
charge in the background of four-dimensional GB BH. Our focus is to
investigate the photon rings and lensing rings surrounding the BH
shadow and observed the light intensity by the thin accretion disk.

\subsection{Direct Emission, Lensing Emission and Photon Ring Emission}

To differentiate the photon ring and lensing ring near the BH, one
can define the total number of light trajectories by
$n(b)=\varphi/2\pi$, where $\varphi$ represents the total change in
azimuthal angle beyond the horizon \cite{12}. The optical appearance
depends upon how near the impact parameter $b$, to its critical
curve $b_{c}$. According to \cite{12}, the trajectories of photon
orbits are mainly divided into three cases and depending upon the
number of orbits around the BH solution such as, $n<3/4$ corresponds
to direct emission, the light trajectories crossing the equatorial
plane just once, while the lensing ring corresponds to $3/4<n<5/4$,
the light trajectories crossing the equatorial plane at leat $2$
times and when $n>5/4$ it corresponds to photon ring, the light
trajectories crossing the equatorial plane minimum $3$ times.

From Fig. \textbf{4}, one can see the that classify regions of
direct, lensing and photon ring emissions for each set of
parameters. We take different values of parameters for each set for
example, set $1$, $2$ and $3$ correspond to $Q=0.3, \phi=0.05,
a=0.1, \alpha=-5.55$, $Q=0.1, \phi=0.01, a=0.1, \alpha=-2.55$ and
$Q=0.2, \phi=0.01, a=0.2, \alpha=0.22$, respectively. The intervals
of $b$ which is related to numerical regions of emissions are listed
as
\begin{equation}\nonumber
\text{Set}:1
    \begin{cases}
    & \text{Direct emission: $n<3/4, b<6.78545$ and $b>8.13506$},\\
    & \text{Lensing
ring: $3/4<n<5/4$, $6.78545<b<6.8203$ and}\\& \text{$6.82030<b<8.13506$},\\
      & \text{Photon
ring: $n>5/4$, $6.82737<b<6.83069$},
    \end{cases}
\end{equation}
\begin{equation}\nonumber
\text{Set}:2
    \begin{cases}
    & \text{Direct emission: $n<3/4, b<6.48387$ and $b>7.64494$},\\
    & \text{Lensing
ring: $3/4<n<5/4$, $6.48387<b<6.55920$ and}\\& \text{$6.57435<b<7.59970$},\\
      & \text{Photon
ring: $n>5/4$, $6.55920<b<6.57435$},
    \end{cases}
\end{equation}
\begin{equation}\nonumber
\text{Set}:3
    \begin{cases}
    & \text{Direct emission: $n<3/4, b<5.85057$ and $b>8.98695$},\\
    & \text{Lensing
ring: $3/4<n<5/4$, $5.85057<b<6.333309$ and}\\& \text{$6.51403<b<8.98695$},\\
      & \text{Photon
ring: $n>5/4$, $6.33330<b<6.51403$},
    \end{cases}
\end{equation}
\begin{figure}[thpb]\centering
\includegraphics[width=5.9cm,height=4.6cm]{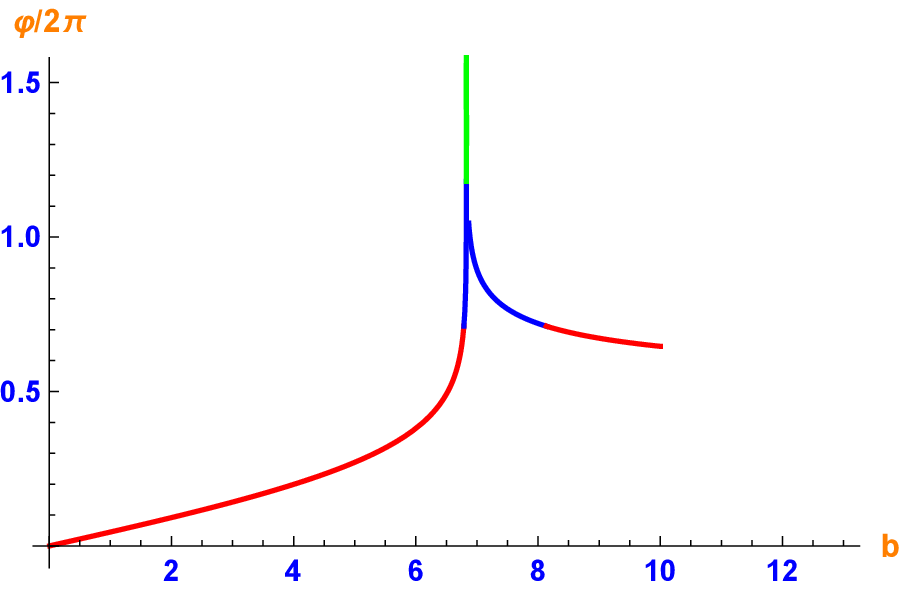}
\includegraphics[width=5.9cm,height=4.6cm]{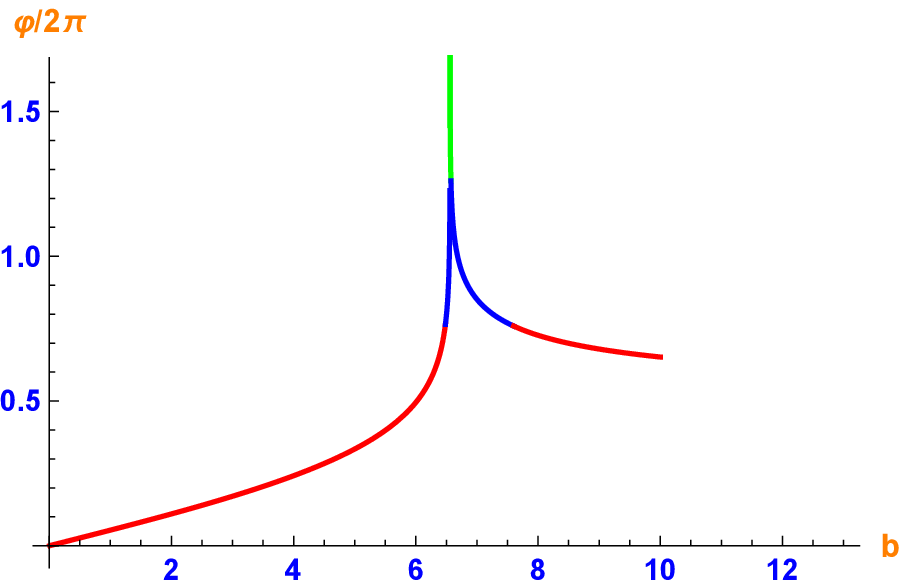}
\includegraphics[width=5.9cm,height=4.6cm]{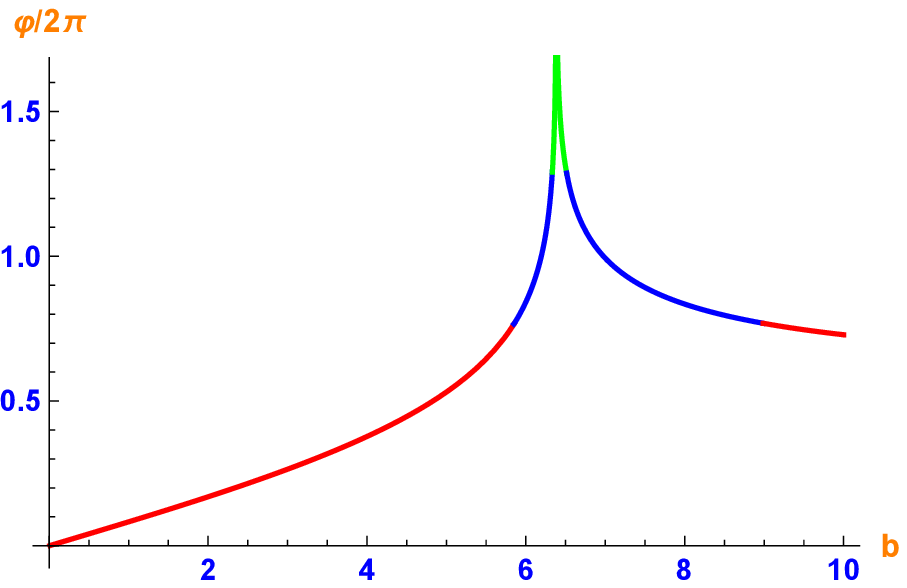}
\includegraphics[width=5.9cm,height=5.6cm]{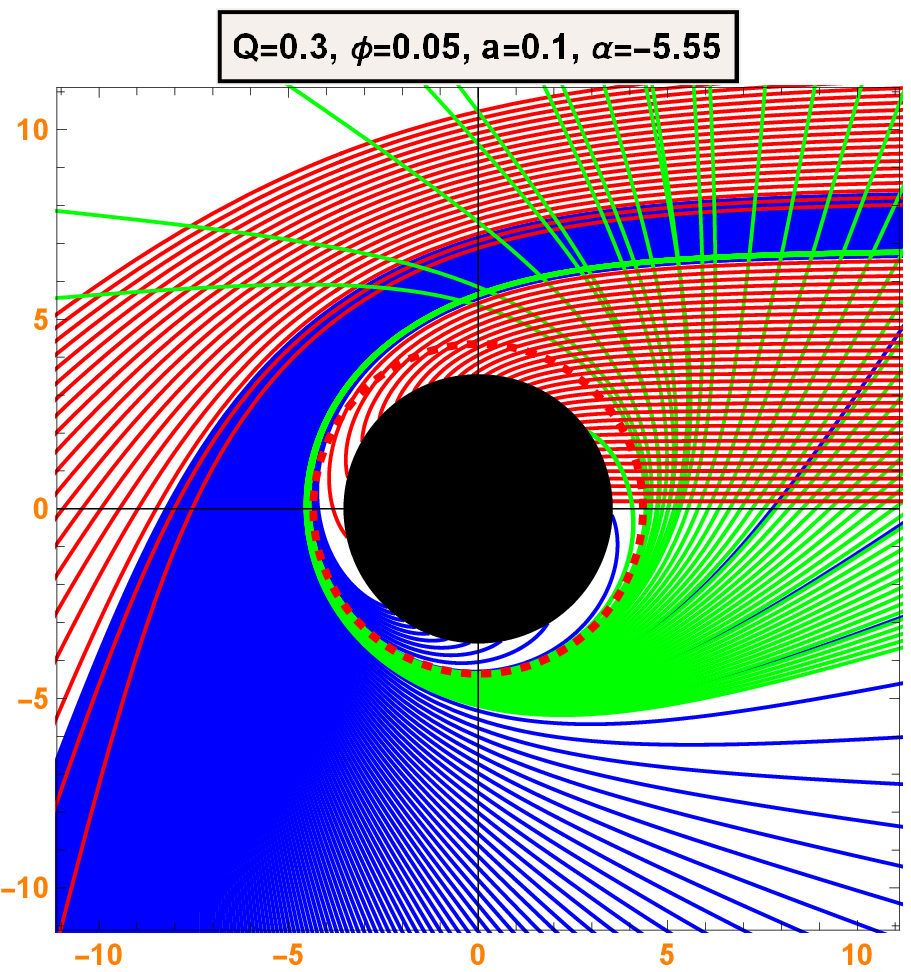}
\includegraphics[width=5.9cm,height=5.6cm]{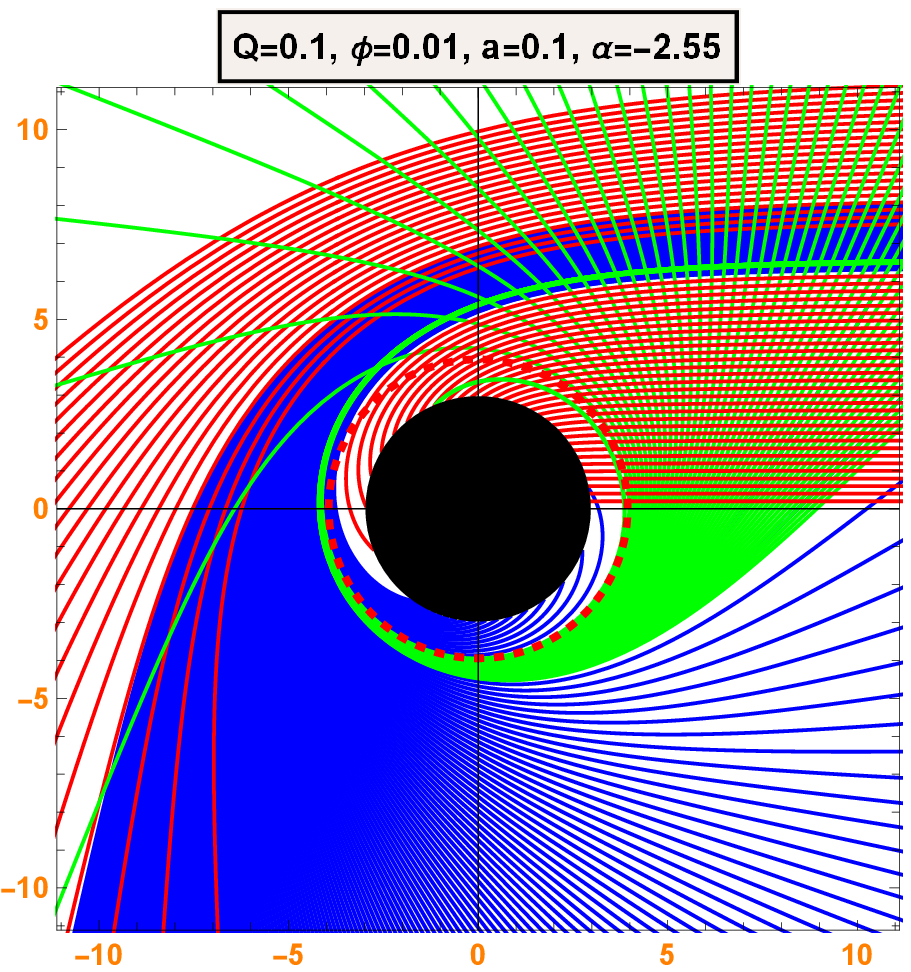}
\includegraphics[width=5.9cm,height=5.6cm]{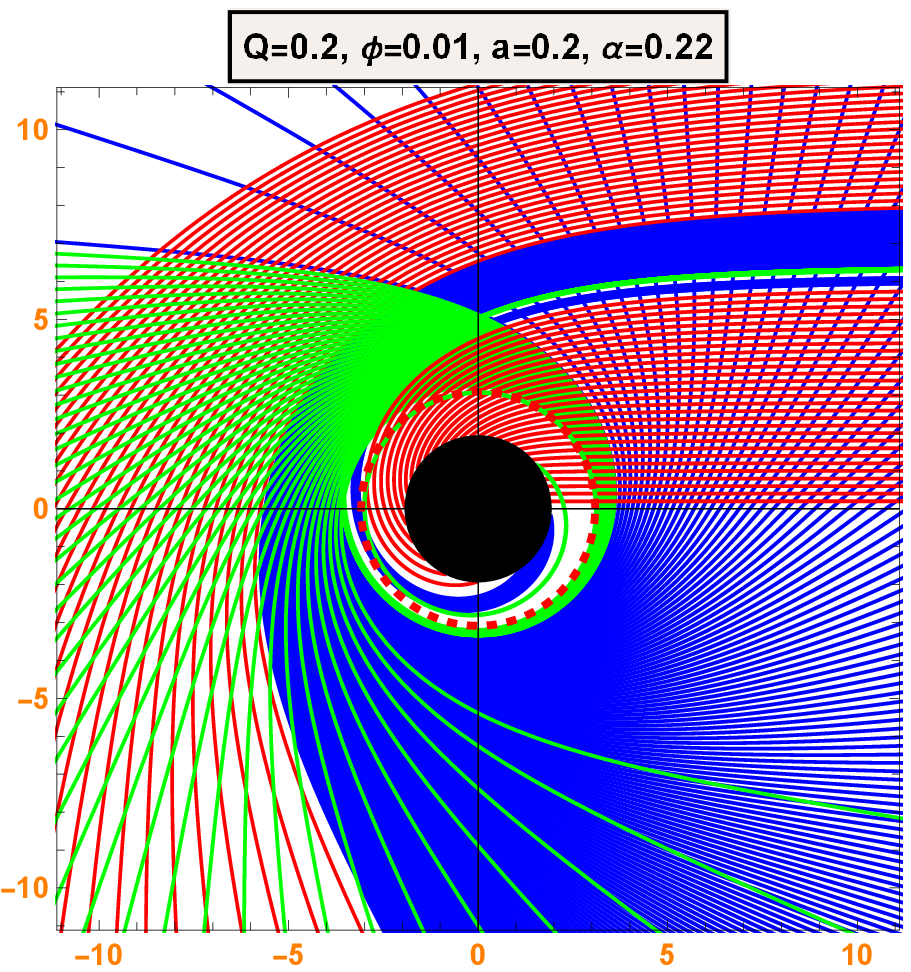}
\caption{The relationship between different trajectories of photons
and impact parameter $b$, for specific values of model parameters in
polar coordinates $(r,\varphi)$. Here, the spacings of $b$ are
$1/5$, $1/100$ and $1/1000$ for direct emission, lensing and photon
ring corresponds to red, blue and green trajectories, respectively.
The BH is shown as solid black disk and dashed red line represents
the photon orbit and the mass of BH as $M=1$.}
\end{figure}
It is worth mentioning that the physical interpretation of photon
behavior around the BH is different for each set. With the variation
of numerical values, the radius of BH is gradually varying and
hence, the bands of photon, lensing, and direct emissions are also
changing the brightness and path of trajectories.

\subsection{Observational Appearance and Transfer Functions}

Now, we are going to analysis the profile of a BH with an
optically/geometrically thin disk accretion and to observe the
specific intensity. We consider that the disk is in the rest frame
of static world-lines, and the emission of photons from it should
obey the fundamental features of isotropy. In addition, as argue in
\cite{12}, we consider that the viewer is static and located at the
zone of north pole and ignored the influence of other lights emitted
from different sources in space-time and only focus on the light
intensity which is emitted from a thin disk.

In this scenario, we delegate the specific emitted intensity and
frequency of the accretion disk as $I_{\text{e}}(r)$ and
$\nu_{\text{e}}$, where observed specific intensity and frequency
are denoted as $I_{\text{obs}}(r)$ and $\nu$. Using the fact of
Liouville's theorem, the quantity
$I_{\text{e}}(r)/\nu_{\text{e}}^{3}$ is conserved in the direction
of light propagation and hence the observed specific intensity can
be defined as
\begin{equation}\label{20}
I_{\text{obs}}(r)=f(r)^{\frac{3}{2}}I_{\text{e}}(r).
\end{equation}
The total observed intensity is deduced by integration over the
entire range of different frequencies as
\begin{eqnarray}\label{21}
I=\int I_{\text{obs}}(r)d\nu =\int
f(r)^{2}I_{\text{e}}d\nu_{\text{e}} =f(r)^{2} I_{\text{em}}(r),
\end{eqnarray}
where $I_{\text{em}}(r)=\int I_{\text{e}}d\nu_{\text{e}}$ is the
total emitted radiation intensity near the thin accretion. From the
previous discussion, if any photon light ray traced backwards from
the observers screen passes through the thin disk accretion plane
once (see blue and green lines from Fig. \textbf{4}), it will get
more light rings from the emission disk .

Hence, the total received optical luminosity will be the sum of all
the intensities from each intersection, mathematically defined as
$I(r)= \sum\limits_{\substack{p}}
f(r)^{2}I_{\text{em}}|_{r}=r_{p}(b)$, where $r_{p}(b)$ is called the
transfer function, containing the information about the radial
position of the $p ^{th}$ intersection between the light with the
impact parameter $b$ and the disk. Moreover, the slope of the
transfer $dr/db$, represents the demagnification factor of the image
\cite{14}. In Fig. \textbf{5}, the red line corresponds to the
direct emission and represents the \textit{first transfer function}
for $p=1$. Since the profile of the direct image is the red-shift
source profile therefore its slope is almost equal to unity.
\begin{figure}[thpb]\centering
\includegraphics[width=5.9cm,height=4.6cm]{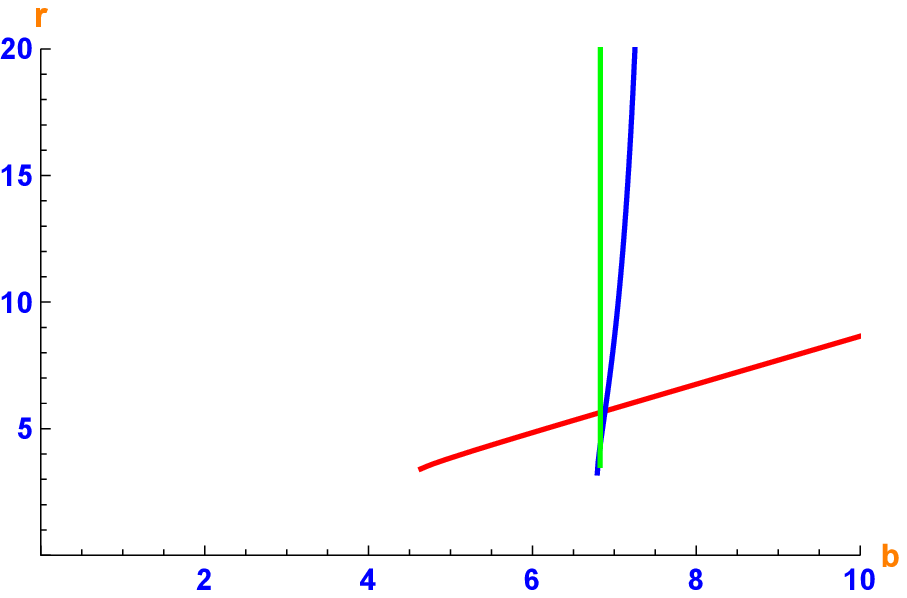}
\includegraphics[width=5.9cm,height=4.6cm]{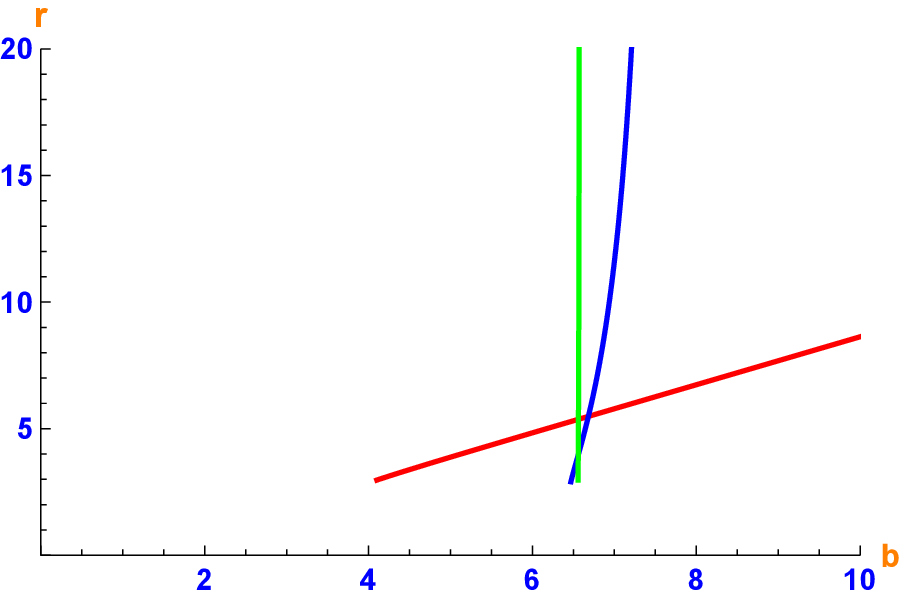}
\includegraphics[width=5.9cm,height=4.6cm]{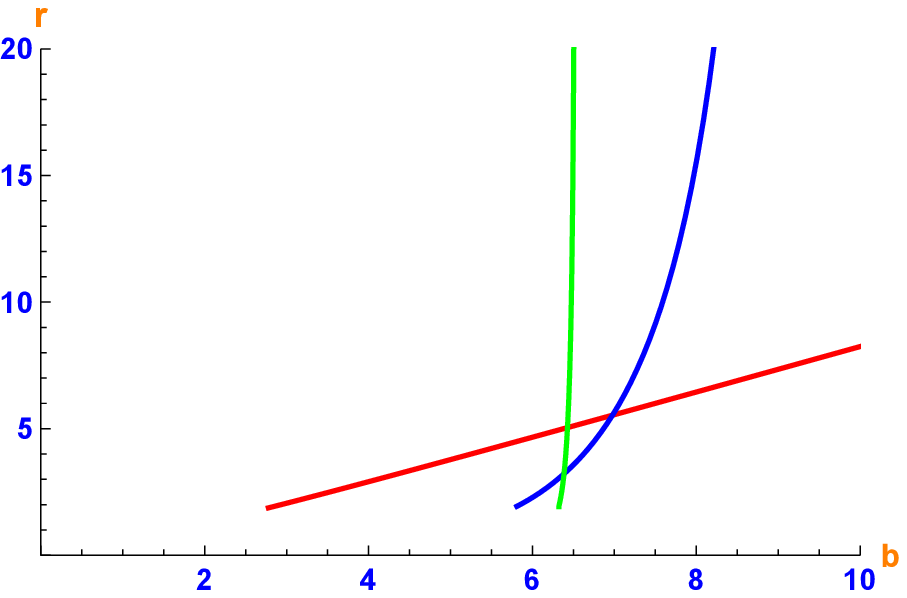} \caption{The profile
depicts the \textit{first three transfer functions} of the BH under
different values of parameters. Here, we plot left to right panels
with the values of set $1$, $2$ and $3$, respectively and the mass
of BH as $M=1$.}
\end{figure}
The blue line corresponds to the lensing ring and represents the
\textit{second transfer function} for $p=2$. Here, the impact
parameter $b$, is closed to the critical curve $b\sim b_{c}$. One
can see that as the value (behaviour) of impact parameter $b$
increases, the slope of the \textit{second transfer function} also
increases, and hence, the back side appearance of the thin disk will
be (de)magnified due to its large slope. The green line corresponds
to the photon ring and represents the \textit{third transfer
function} for $p=3$. The slope $dr/dp$ is closed to $\infty$ so, the
appearance of the front side of the thin disk will be highly
demagnified. Further, we ignored the later transfer functions safely
because the image depicted from these transfer functions is
extremely (de)magnified and negligible.

\subsection{Specific Luminosities of Thin Accretion Disks}

Now, we consider an optically and geometrically thin disk model to
observe the further specific intensity surrounding our BH solution
on the equatorial plane. (Actually, this type of model evaluate the
accretion matters when the accretion flow velocities is
sub-Eddington with large opaque but ignores those with high
accretion velocities and mass \cite{32}. For instance, around the
supermassive BHs, the accretion disk may effectively turn into an
apparently thin but structurally thick one \cite{33}). The main
source of observed specific light intensity is a thin disk and its
luminosity only depends upon the radial coordinate $r$. So, we
assume the following three toy models which is used to evaluate the
realistic cases of thin matters.
\begin{itemize}
\item Model $1$: We assume that the matter emission begins from the peak
point of the radius at the inner-most stable circular orbit (isco)
for time-like observers. Therefore, we consider the model for this
emission profile is a decay function, which is defined as
\begin{equation}\label{22}
I^{1}_{\text{em}}(r)=
    \begin{cases}
     \text{$(\frac{1}{r-(r_{\text{isco}}-1)})^{2},$ \quad if $r>r_{\text{isco}}$} \\
     \text{$0$,\quad \quad \quad \quad \quad \quad if $r\leq r_{\text{isco}}$}
    \end{cases}
\end{equation}
\begin{figure}[thpb]\centering
\includegraphics[width=5.9cm,height=4.6cm]{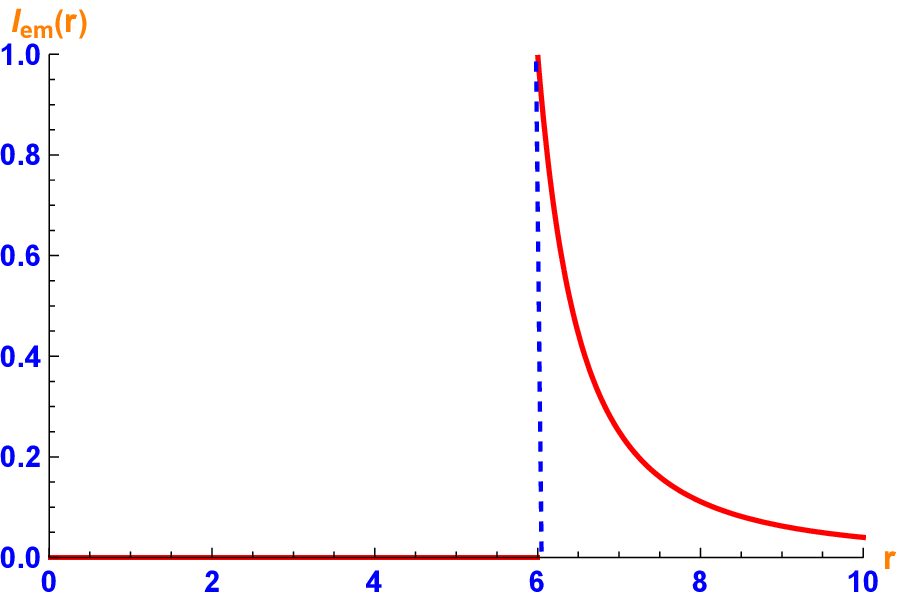}
\includegraphics[width=5.9cm,height=4.6cm]{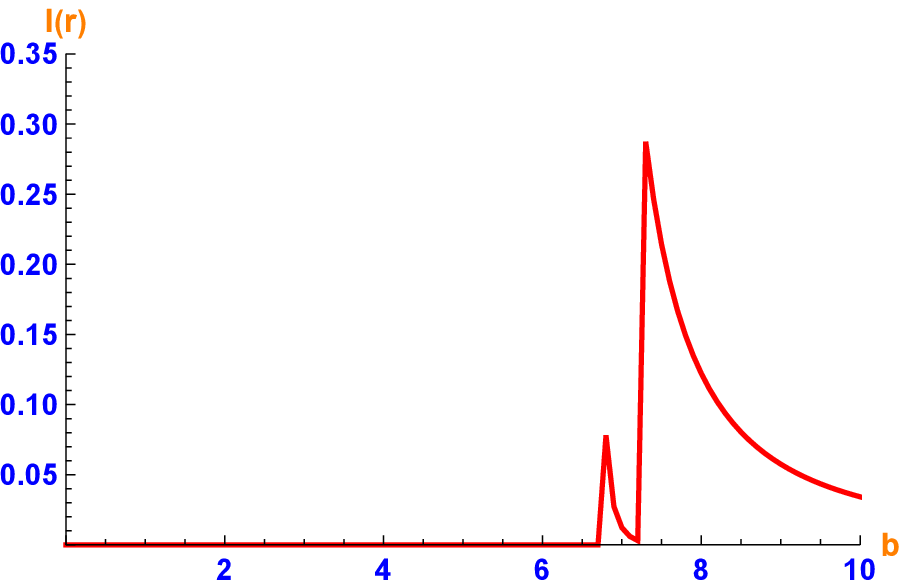}
\includegraphics[width=5.9cm,height=4.6cm]{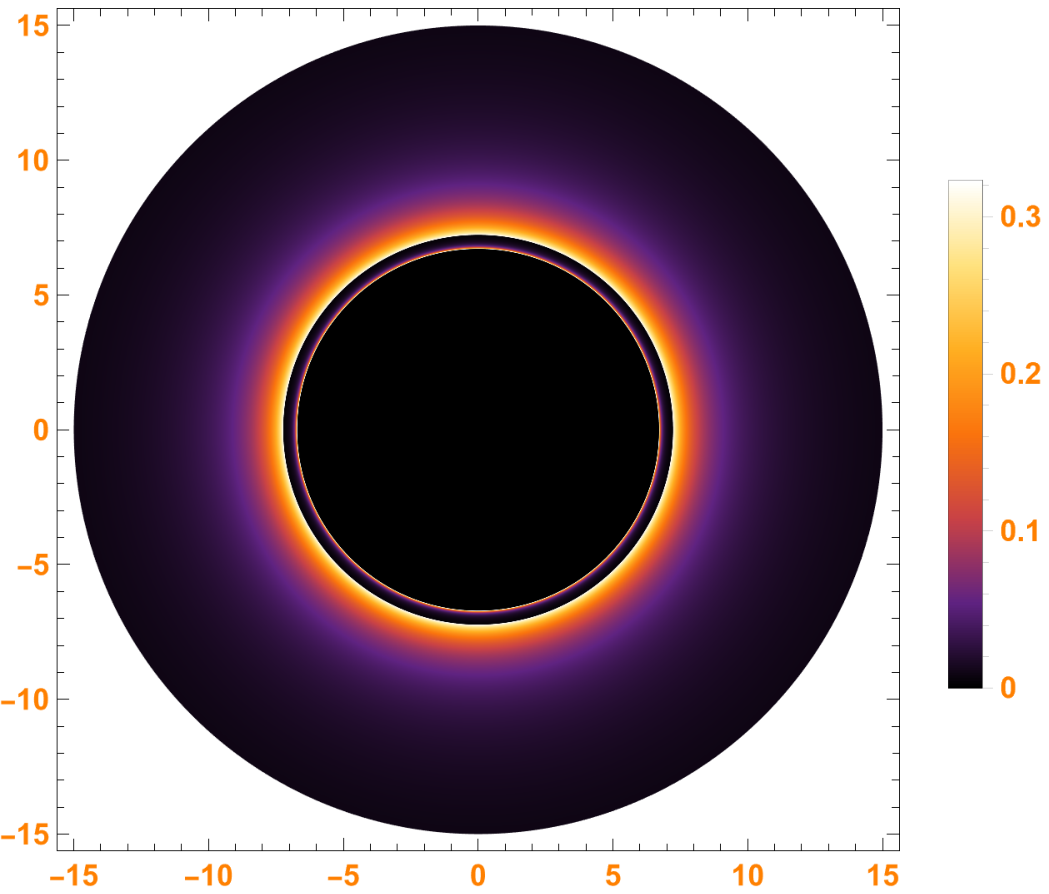}
\includegraphics[width=5.9cm,height=4.6cm]{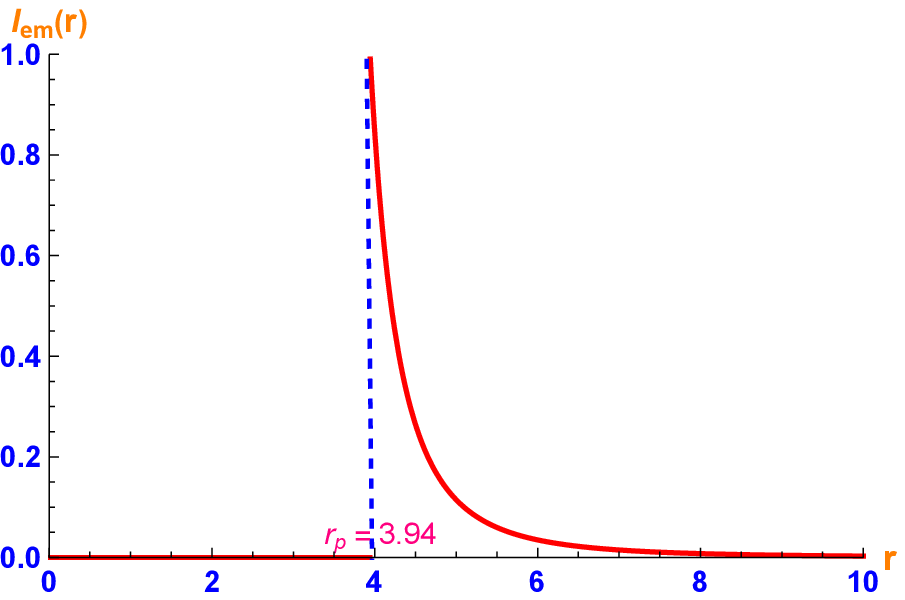}
\includegraphics[width=5.9cm,height=4.6cm]{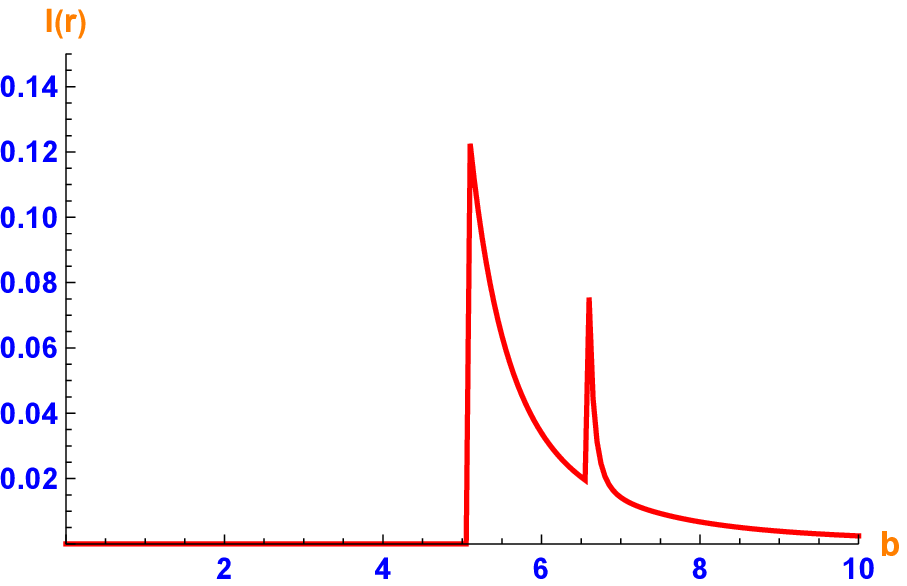}
\includegraphics[width=5.9cm,height=4.6cm]{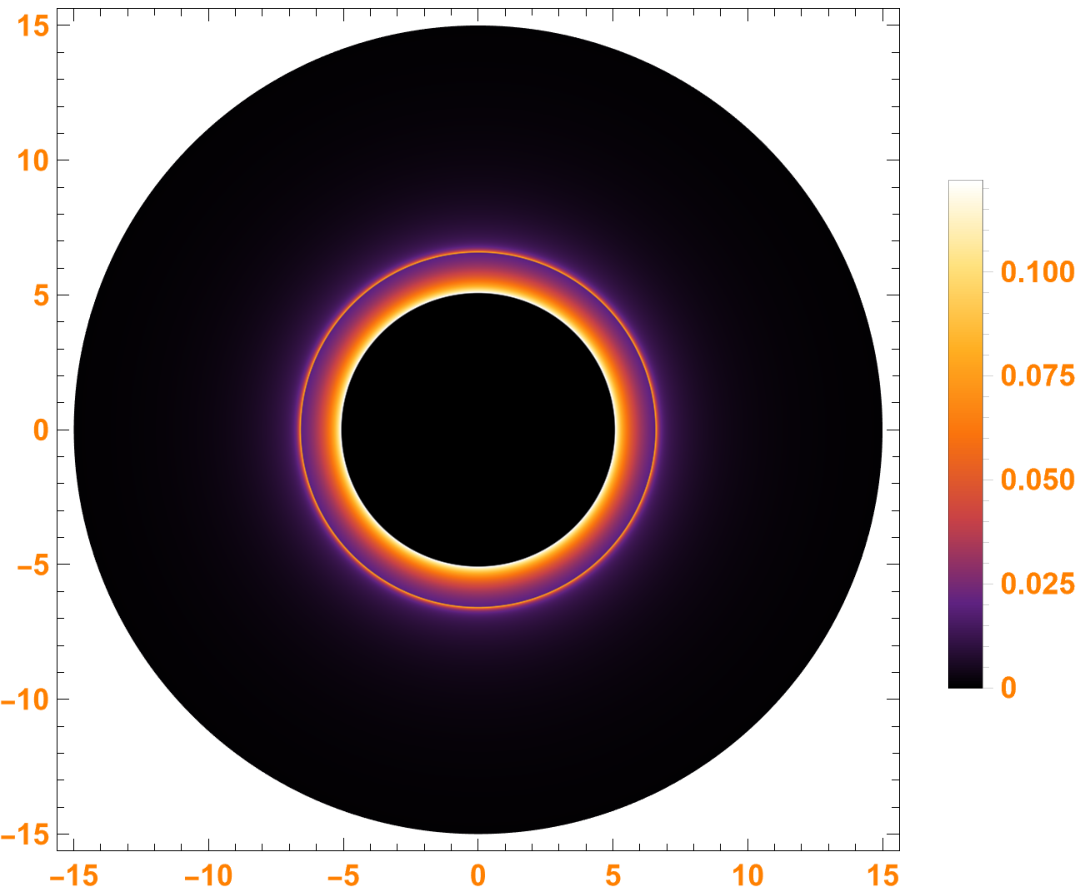}
\includegraphics[width=5.9cm,height=4.6cm]{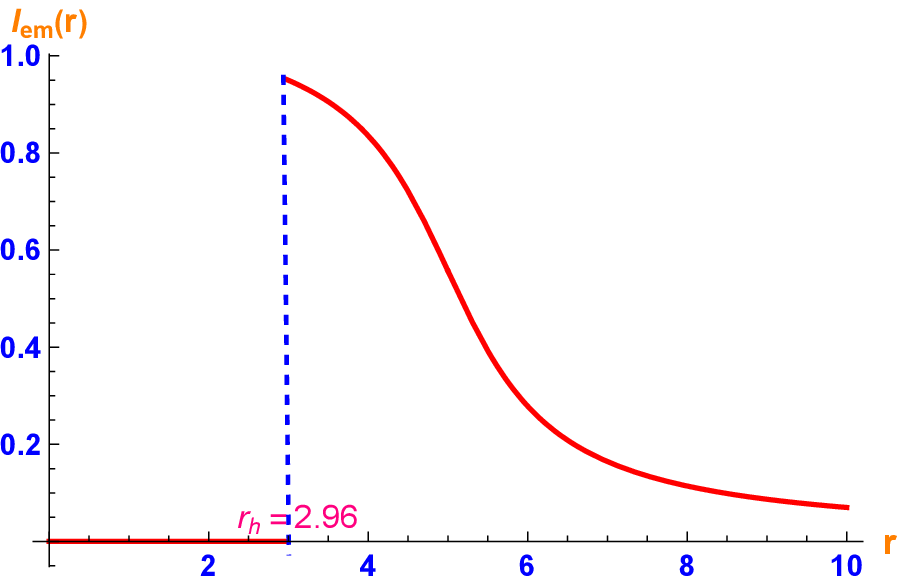}
\includegraphics[width=5.9cm,height=4.6cm]{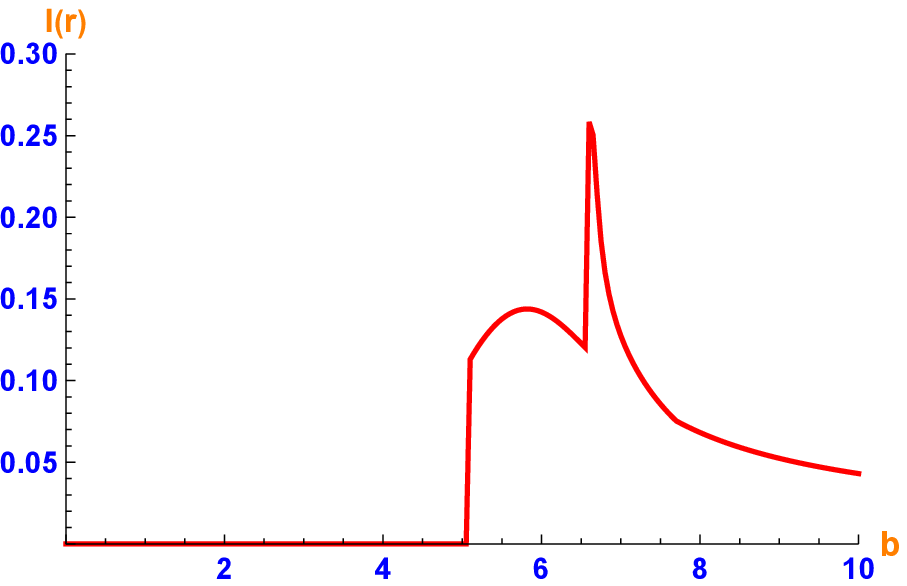}
\includegraphics[width=5.9cm,height=4.6cm]{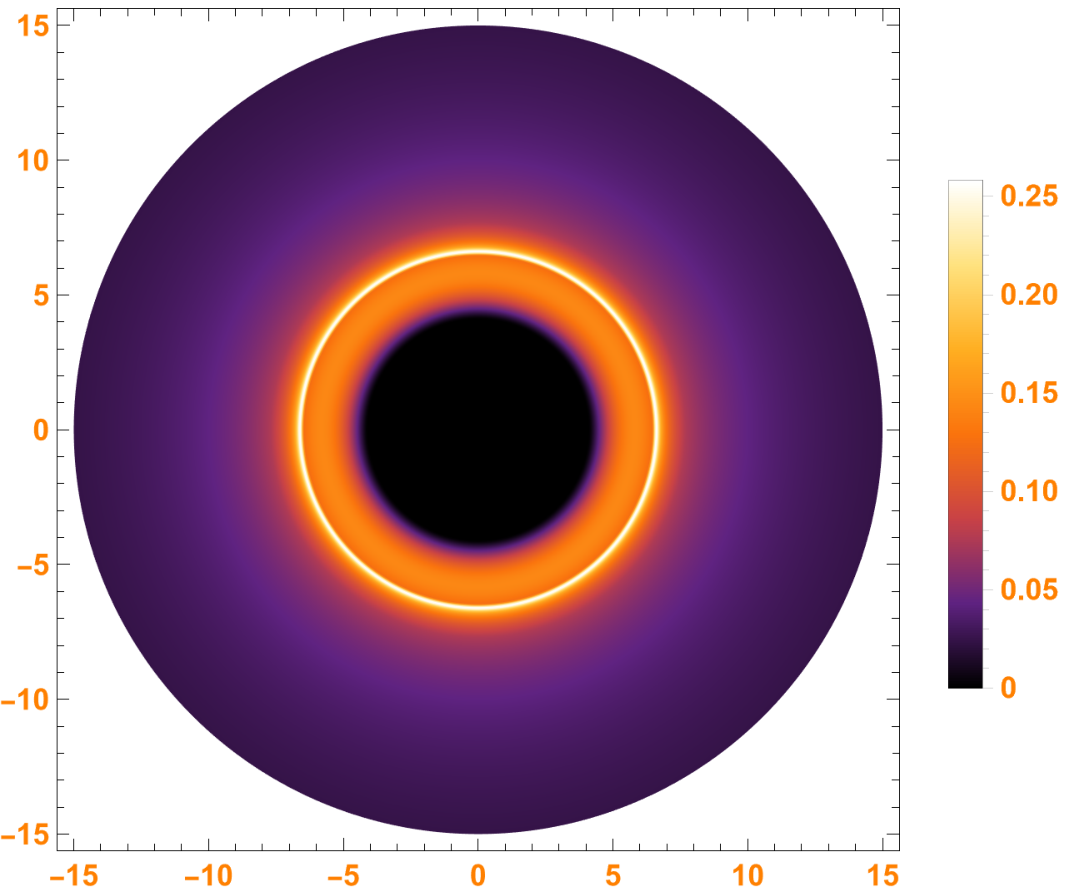}
\caption{Optical observation of thin disk accretion near the BH with
different emission matter profiles, viewed from a face-on
orientation. From left to right panels, the depicted graphs are
reflecting the various emissions, observed emission and two
dimensional optical appearance, respectively. The numerical values
of parameters for all profiles are defined in set $2$ and the mass
of BH as $M=1$.}
\end{figure}
\item Model $2$: the considering emission has a sharp spikes at the isco location, having relatively similar
center and asymptotic dynamics as model $1$. But the emission
luminosity attenuation is significantly larger, so that the emission
has decay characteristics of the third-order, mathematically defined
as
\begin{equation}\label{23}
I^{2}_{\text{em}}(r)=
    \begin{cases}
     \text{$(\frac{1}{r-(r_{p}-1)})^{3},$ \quad if $r>r_{p}$} \\
     \text{$0$,\quad \quad \quad \quad \quad if $r\leq r_{p}$}
    \end{cases}
\end{equation}
\item Model $3$: This emission lie beyond the horizon $r_{h}$, but
its decaying rate is moderate than the previous two cases, as
defined below
\begin{equation}\label{24}
I^{3}_{\text{em}}(r)=
    \begin{cases}
     \text{$\frac{\frac{\pi}{2}-\arctan(r-5)}{\frac{\pi}{2}-\arctan(-3)},$ \quad if $r>r_{h}$} \\
     \text{$0$,\quad \quad \quad \quad \quad if $r\leq r_{h}$}
    \end{cases}
\end{equation}
\end{itemize}
\begin{figure}[thpb]\centering
\includegraphics[width=5.9cm,height=4.6cm]{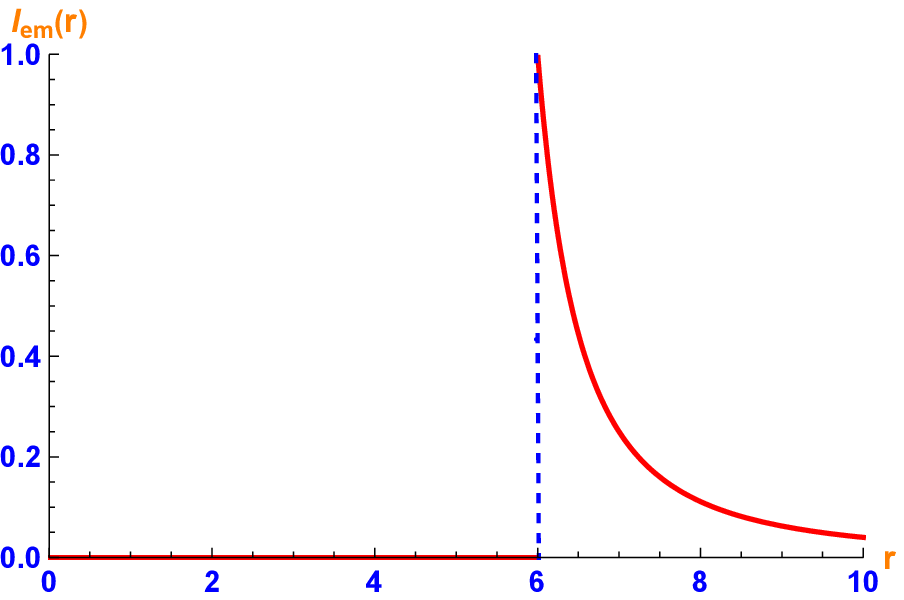}
\includegraphics[width=5.9cm,height=4.6cm]{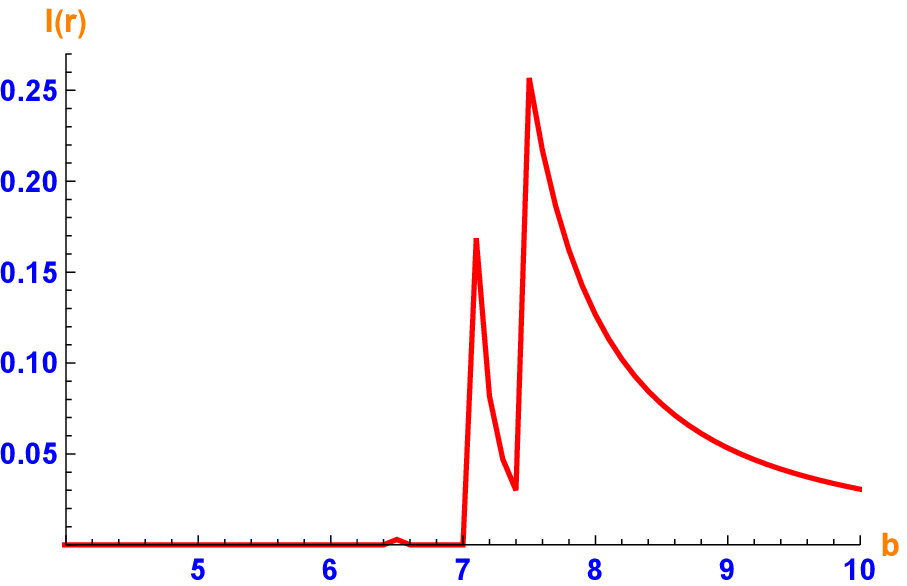}
\includegraphics[width=5.9cm,height=4.6cm]{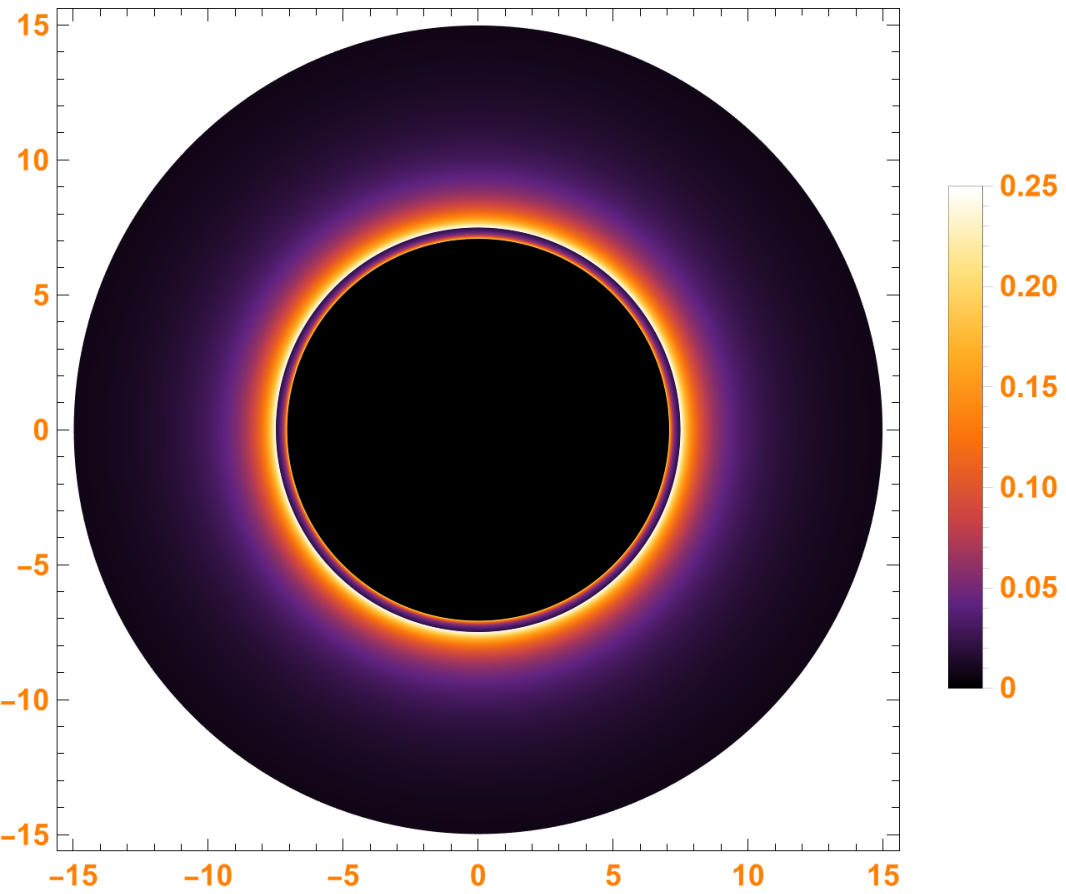}
\includegraphics[width=5.9cm,height=4.6cm]{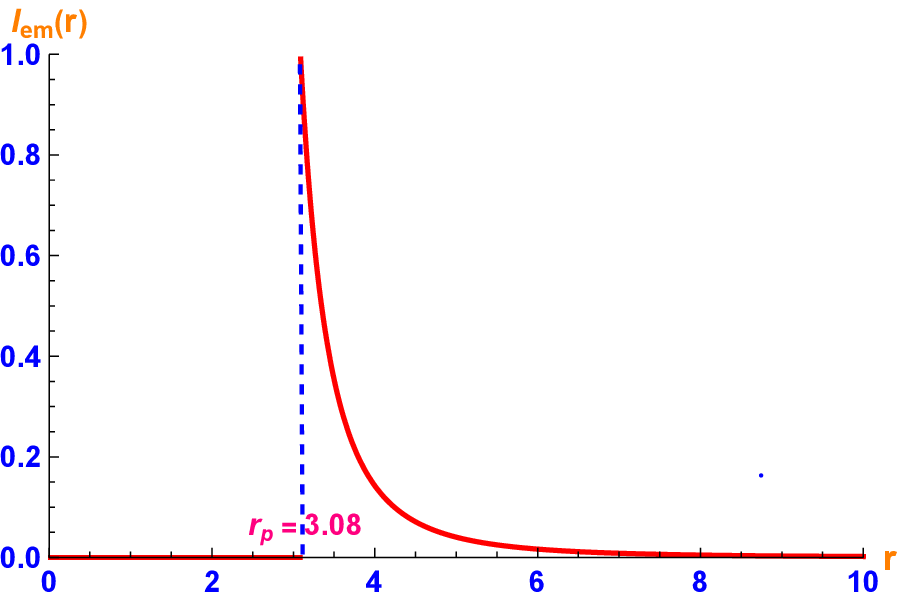}
\includegraphics[width=5.9cm,height=4.6cm]{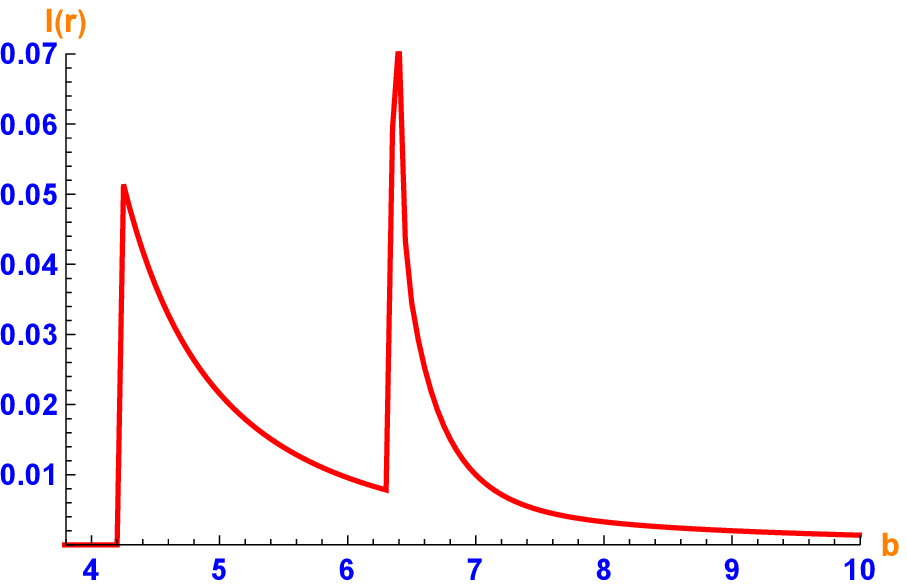}
\includegraphics[width=5.9cm,height=4.6cm]{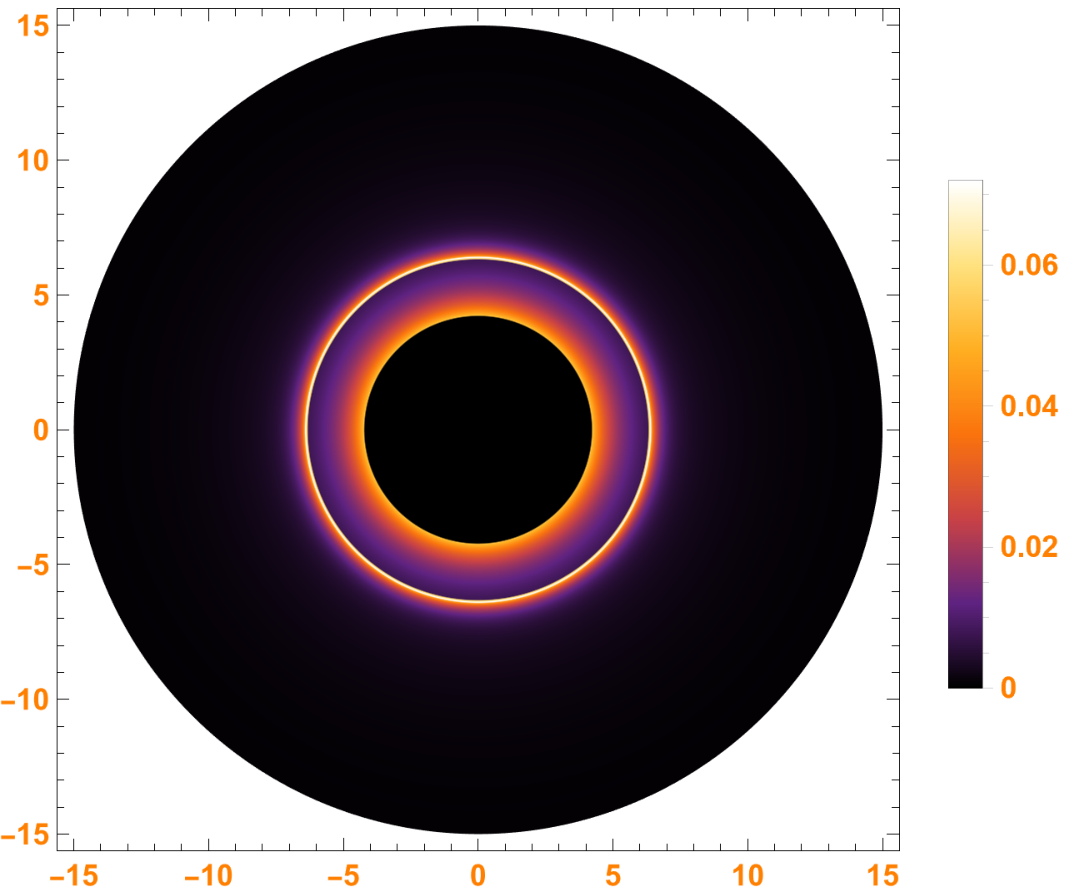}
\includegraphics[width=5.9cm,height=4.6cm]{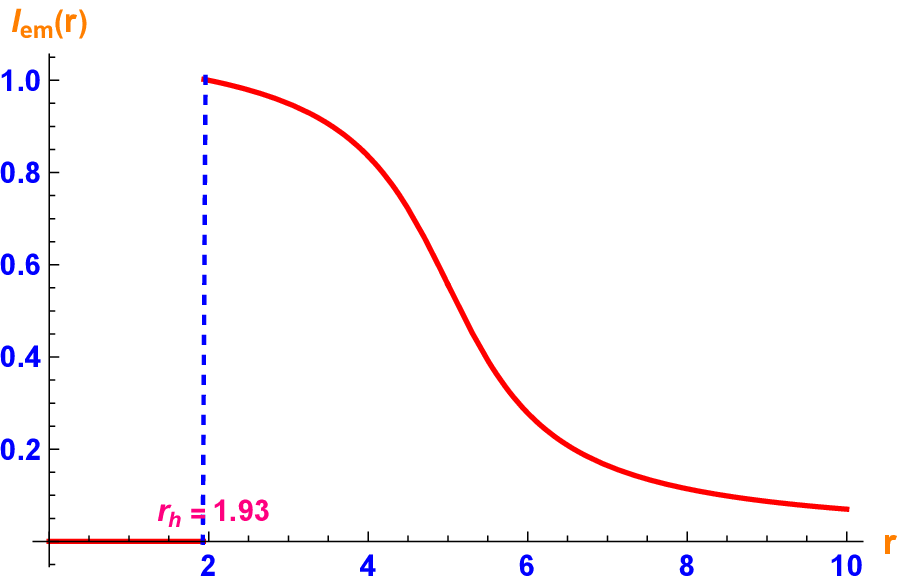}
\includegraphics[width=5.9cm,height=4.6cm]{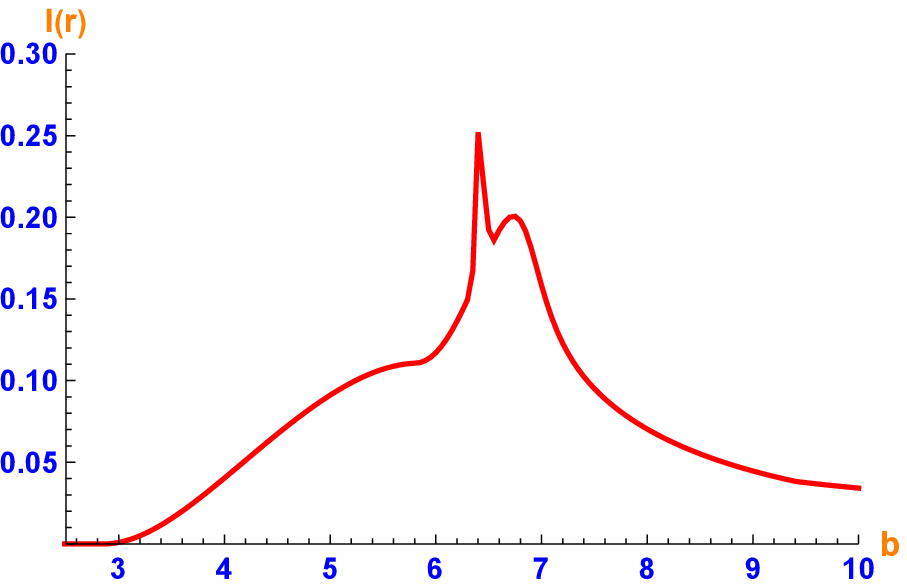}
\includegraphics[width=5.9cm,height=4.6cm]{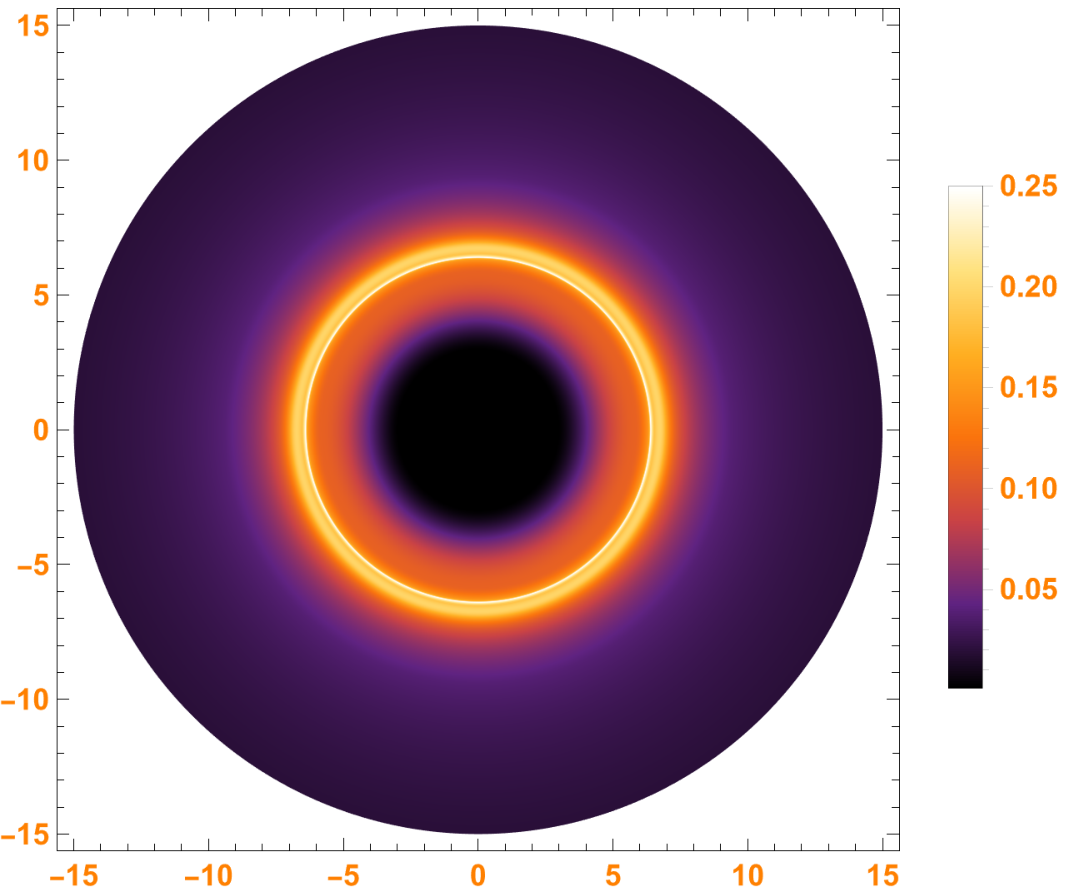}
\caption{Optical observation of thin disk accretion near the BH with
different emission matter profiles, viewed from a face-on
orientation. From left to right panels, the depicted graphs are
reflecting the various emissions, observed emission and two
dimensional optical appearance, respectively. The numerical values
of parameters for all profiles are defined in set $3$ and the mass
of BH as $M=1$.}
\end{figure}

With the help of previously defined toy models, we consider just two
examples for different numerical values of parameters as defined in
Figs. \textbf{6} and \textbf{7}, and showed the results of intensity
and analyzed the emission under these constraints. In Figs.
\textbf{6} and \textbf{7}, the left column shows the emission
profile from the accretion has good behavior outside the photon
sphere and defines a connection between specific emission intensity
and the radial coordinate $r$. The middle column depicts the one
dimensional (observed intensity) function of $I(r)$, which is
related to $b$, where as the right column is the two dimensional
density profiles that reflect the optical appearance of observed
intensities.

From Fig. \textbf{6} (first row, left panel), we observed that the
emission flow has reached its peak value, close to the critical case
of the impact parameter $r\sim b$, after that decaying dramatically
to zero with the increasing values of $r$. In the middle panel, due
to the intensity of gravitational lensing, we see that there are two
spikes separated the photon and lensing orbit independently, and the
corresponding observed image decays similar to $b>7.48M$. However,
the confined region of the lensing ring lies in the range
$6.85M<b<7.34M$, providing only a small contribution to the total
flux while the photon ring contribute the negligible luminosity.
Therefore, one needs to see the main contribution in the observed
(optical) appearance which is obtained from the direct emission and
yields a wide rim while lensing ring makes a small contribution, and
we find it inner of a wide rim. Moreover, the innermost region
represents the photon ring contribution to the total flux which is
difficult to detect in the right panel of density profiles and
spikes at $b\sim6.64M$.

From the second row of Fig. \textbf{6}, the emission attained a peak
value near the photon sphere at $r_{p}=3.94$ as shown in the left
panel. The view of the middle panel reflects that the direct
emission intensity shows the maximum value at $b=5.11$ and then
shows a nice fluctuating behavior with an increase of $b$. Further,
the region of lensing ring makes a significant intensity in the wide
range of $b$ lie in $6.55M<b<6.83M$, while the photon ring emission
has a narrower the spike at $b\sim6.64M$ which can hardly be
differentiated from the lensing one which leads to highly
demagnetized to a narrow region and the intensity of the direct
emission is still observed dominant. The overall results in the two
dimensional optical appearance is visualized in the right panel. The
lensing ring has small contribution to the total brightness while
the photon ring is hardly diluted and narrowly visible.

Finally, in the third row of Fig. \textbf{6}, the emitted region has
been increased to the event horizon $r_{h}=2.96$, as shown in the
left column and the decay rate of the emission is very moderate as
compared to previous two models. One can see that from the inner
rim, the observed intensity lies in the lensed position of the event
horizon at $b\sim0.25M$. The observational intensity increases
suddenly and reaches at highest point in the emission of the photon
ring and lies beyond the dark region due to the influence of
gravitational red-shift. After this, the observed intensity start to
show decaying behaviour nicely at $b\sim6.64M$ in the photon ring
and the participation of the lensing ring to the total flux is
appreciable as compared to previous two cases. In this case, the
observed appearance reflect a narrow but makes a prominent brighter
extended ring contribution to the observed intensity, but the photon
ring is still safely ignored.

From Fig. \textbf{6}, it is interesting to mention that the dark
interior regions are different for different plots of emissions, but
the location of the photon ring stays always at $b\sim6.64M$. In
addition, we plot all the emission profiles and observed intensities
in Fig. \textbf{7}. The graphical description is the same for these
profiles as we discussed in Fig. \textbf{6}. The differences are the
location of the photon/lensing rings and the quantities of
luminosity intensities. The optical appearance in these flow models
is physically well behave and viable with the statistical mechanics
obtained from the original analysis of the Schwarzschild BH as
discussed in \cite{12}.

\section{Shadows of the BH with Rest Spherical Accretion}

Here, we are going to analyze the shadows of the BH with static
spherical accretion model for various values of model parameters. We
investigate the shadow of charged four-dimensional GB BH with the
influence of NC parameter and cloud of strings for static spherical
accretion. To this end, we focus on the observed specific intensity
(which is usually defined in
$\text{ergs}^{-1}\text{cm}^{-2}\text{str}^{-1}\text{Hz}^{-1}$), as
expressed in \cite{34,35}.
\begin{equation}\label{25}
I^{\text{obs}}(b)=\int g^{\text{obs}3}j(\nu^{\text{obs}}_{e})dl_{p},
\end{equation}
where $g^{\text{obs}}=\nu^{\text{obs}}_{o}/\nu^{\text{obs}}_{e}$ is
the red-shift factor, $\nu^{\text{obs}}_{o}$ is the observed photon
frequency, $\nu^{\text{obs}}_{e}$ is radiate photon frequency,
$j(\nu^{\text{obs}}_{e})$ represents the emissivity per unit volume
which is calculated in the static frame of the emitter and $dl_{p}$
is the infinitesimal proper length. From Eq. (\ref{2}), the
red-shift factor $g^{\text{obs}}=f(r)^{\frac{1}{2}}$. We consider
the emission of light radiations is monochromatic, which is
perceived with single constant frequency $\nu_{k}$, i.e.,
\begin{equation}\label{26}
j(\nu^{\text{obs}}_{e})\propto\frac{\delta(\nu_{e}-\nu_{k})}{r^{2}}.
\end{equation}
Further, the emission of light has $1/r^{2}$ radial profile as
defined in \cite{35}, and one can derived the proper length in
space-time structure as following
\begin{eqnarray}\nonumber
dl_{p}&=&\sqrt{(f(r)^{-1}dr^{2}+r^{2}d\varphi^{2})},\\\label{27}&=&\sqrt{f(r)^{-1}+r^{2}\bigg(\frac{d\varphi}{dr}
\bigg)^{2}}dr.
\end{eqnarray}
Using Eqs. (\ref{25})-(\ref{27}), we can obtain the specific
intensity which is observed by the static observer as
\begin{eqnarray}\label{28}
I^{\text{obs}}(b)=\int\frac{f(r)^{3/2}}{r^{2}}\sqrt{f(r)^{-1}+r^{2}\bigg(\frac{d\varphi}{dr}\bigg)^{2}}dr.
\end{eqnarray}
\begin{figure}[thpb]\centering
\includegraphics[width=5.9cm,height=4.6cm]{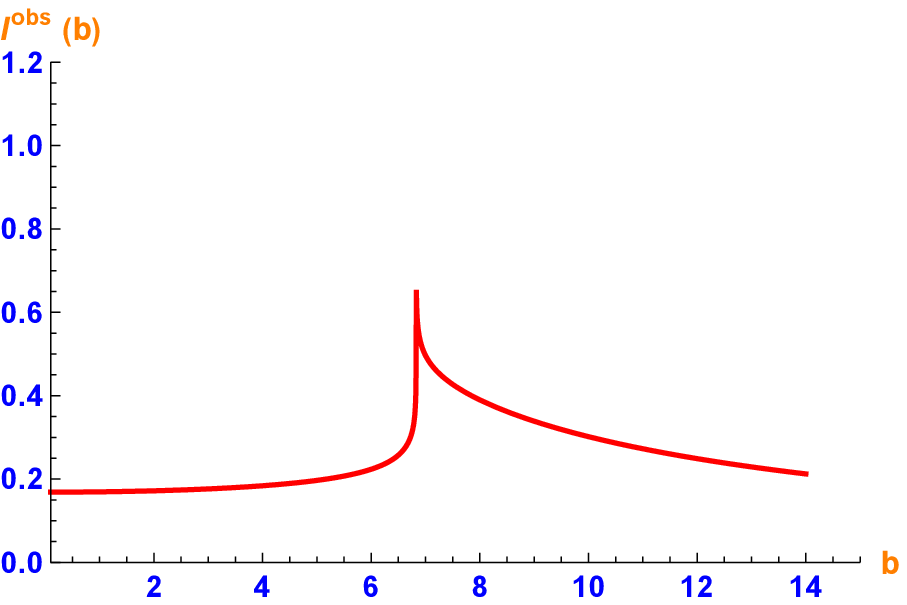}
\includegraphics[width=5.9cm,height=4.6cm]{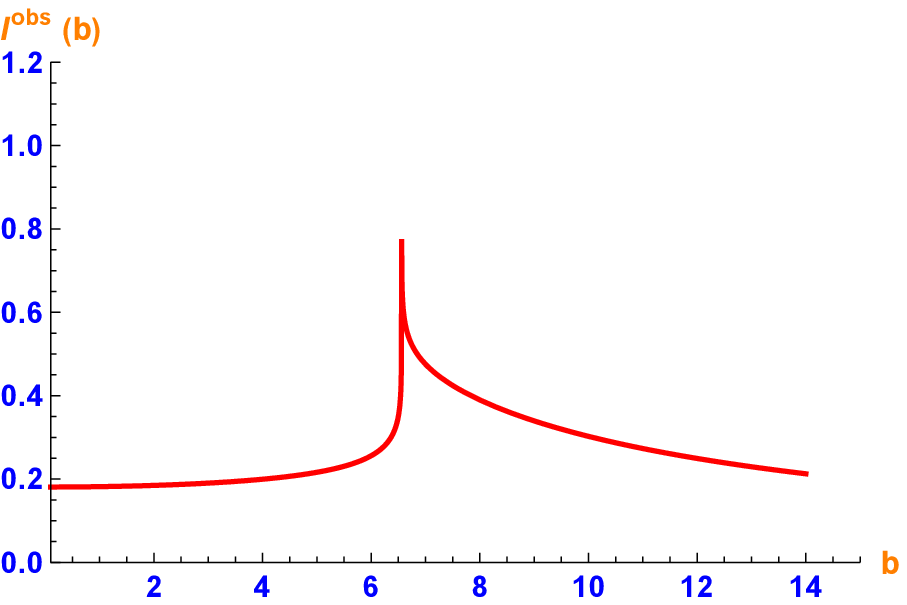}
\includegraphics[width=5.9cm,height=4.6cm]{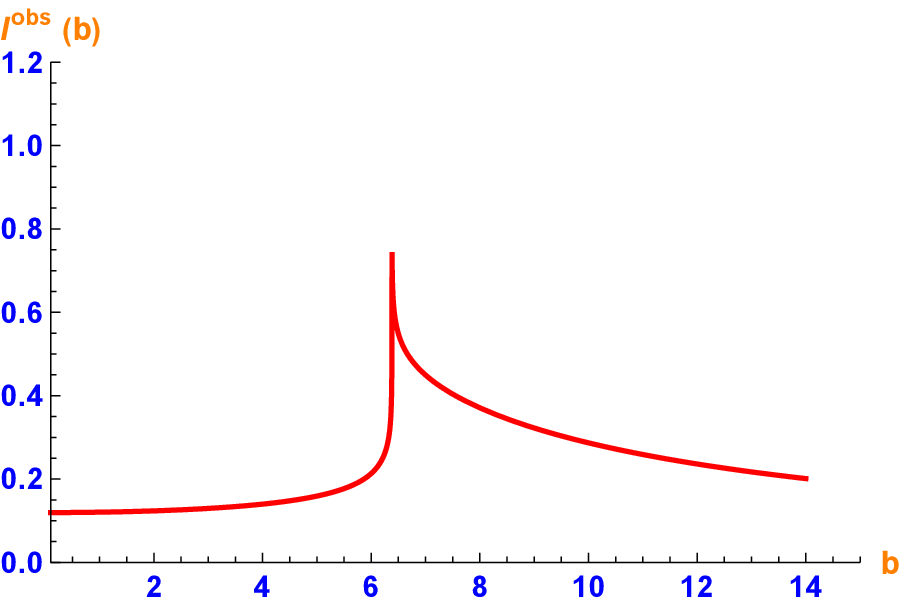}
\includegraphics[width=5.9cm,height=5.5cm]{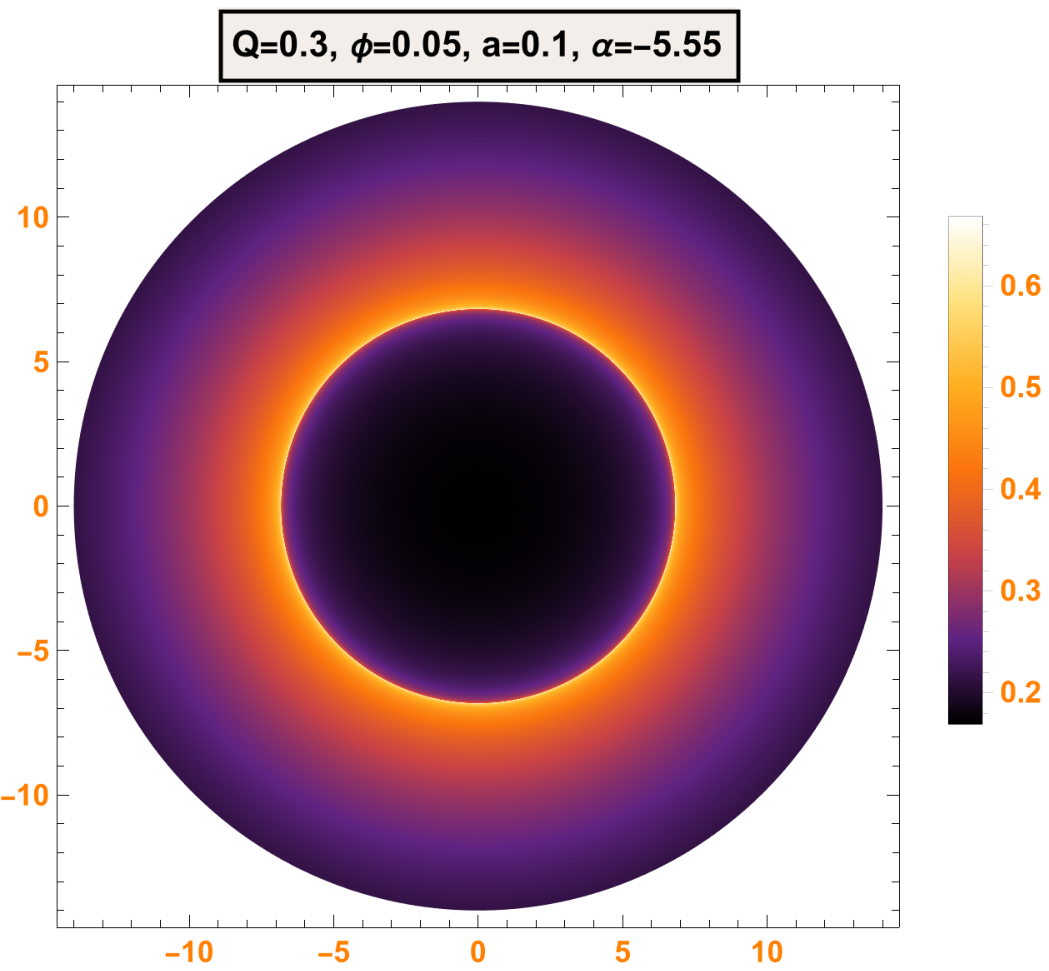}
\includegraphics[width=5.9cm,height=5.5cm]{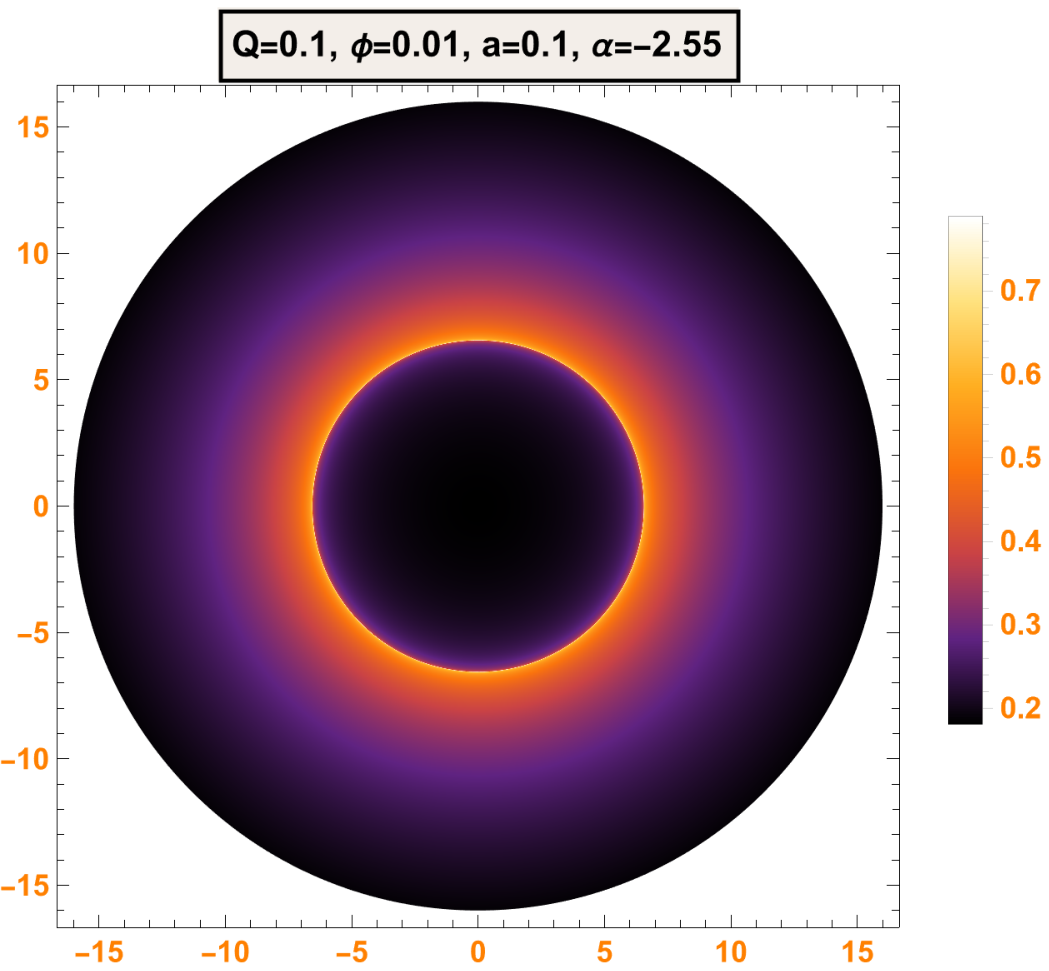}
\includegraphics[width=5.9cm,height=5.5cm]{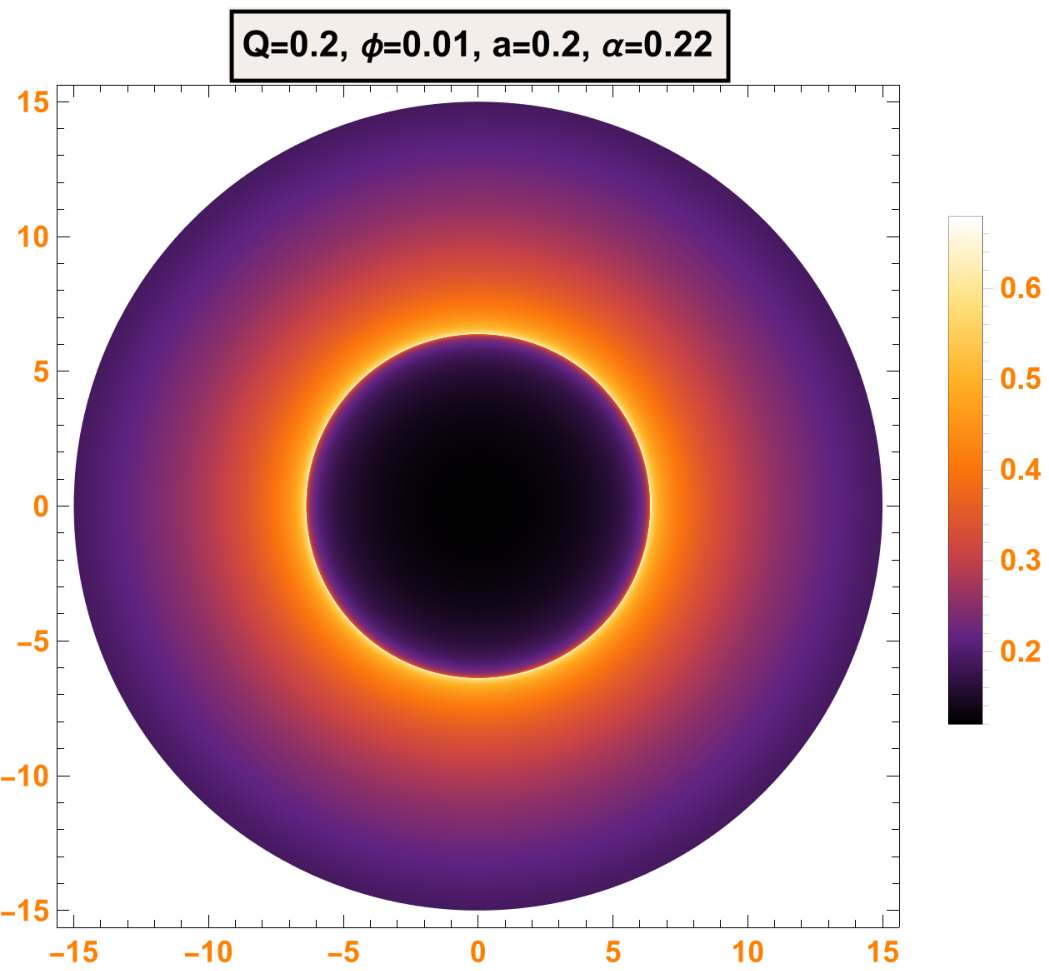}
\caption{The graphical interpretation of the total observed
intensity $I^{\text{obs}}(b)$, reflects by a static spherical
accretion flow matter, for different BH state parameters and the
mass of BH as $M=1$. We choose these numerical values as three
examples, and one can find a comprehensive details in the main
texts.}
\end{figure}

Based on Eq. (\ref{28}), we will discuss the shadow appearance and
related observed intensity of the considering BH in the perspective
of static spherical accretion under the influence of parameters. The
intensity of the light can be measured with the help of light pulse,
which is calculated by the impact parameter $b$. In Fig. \textbf{8}
(top row), the observed intensities of light rays are shown for
various values of model parameters as mentioned in the
two-dimensional BH shadow image in the bottom row.

From Fig. \textbf{1}, we see that as the trajectory of impact
parameter $b$ increases, the observed specific intensity starts to
increase nicely and stay at the peak when $b=b_{c}$, and then
exhibits a decaying pattern of attenuation. From Fig. \textbf{8}
(top row), as $b$ increases, the intensity ascended first when
$b<b_{c}$, then reached the peak value at $b=b_{c}$ and finally
dropped down in the region $b>b_{c}$. This result is consistent with
Fig. \textbf{1} and physically viable. When $b$ is smaller than the
critical case $b_{c}$, the observed intensity originating from
accretion matter is absorbed by the BH. For $b=b_{c}$, the
trajectory of the photon is rotating about the BH several times in
the BH photon ring orbit. Therefore, a distant static observer sees
the maximal luminosity at the critical point.

Meanwhile, for $b>b_{c}$, the specific intensity $I^{\text{obs}}(b)$
shows a decaying behavior, and when $b\rightarrow \infty$, the
observed intensity will be zero. Further, it is also observed that
the intensity of the light ray gradually varies (increasing or
decreasing) with the variation of parameters, and hence, each
parameter plays a significant role in the BH luminosity. The
two-dimensional BH shadow image is also reflected in Fig. \textbf{8}
(bottom row), where different bright rings correspond to different
values of the specific intensity. The image of the BH shadow is
circularly symmetric, and BH is surrounded by a bright photon ring,
so-called photon sphere. The numerical evaluation of photon sphere
radius for some specific values of model parameters are listed in
Table. \textbf{1}. Clearly, the features in Fig. \textbf{8} are
physically viable with that in Table. \textbf{1}. Moreover, the
interior of photon sphere does not vanish completely, because there
is little part of radiative gas has escaped from the BH.
\section{Shadows of the BH with an Infalling Spherical Accretion}

The geometrically thin accreting matter is assumed to be more
realistic in nature due to infalling matters. Now Eq. (\ref{28}) is
still useful, but the associated red-shift factor defined as
\begin{equation}\label{31}
g=\frac{\mathcal{K}_{\sigma}u_{0}^{\sigma}}{\mathcal{K}_{\tau}u_{e}^{\tau}},
\end{equation}
in which $\mathcal{K}^{\rho}=\dot{x}_{\rho}$,
$u_{0}^{\rho}=(1,0,0,0)$ and $u_{e}^{\rho}$ represent the
four-velocity components of photon, static observer and accretion
matter respectively, as given by
\begin{eqnarray}\label{32}
u_{e}^{s}=f(r)^{-1}, \quad u_{e}^{v}=-(1-f(r))^{1/2}, \quad
u_{e}^{\theta}=u_{e}^{\varphi}=0.
\end{eqnarray}
Using Eqs. (\ref{12})-(\ref{14}), we obtained the components of
four-velocity. As $\mathcal{K}_{s}=1/b$ is a constant term and
$\mathcal{K}_{v}$ can be inferred from
$\mathcal{K}_{\varpi}\mathcal{K}^{\varpi}=0$, i.e.,
\begin{equation}\label{33}
\frac{\mathcal{K}_{v}}{\mathcal{K}_{s}}=\pm\frac{1}{f(r)}\sqrt{1-\frac{b^{2}f(r)}{r^{2}}},
\end{equation}
where the sign ``$\pm$'' corresponds to the motion of photons moving
towards/away from the BH. Confronting Eq. (\ref{33}), the red-shift
factor given in Eq. (\ref{31}) can be obtained as
\begin{equation}\label{34}
g=\frac{1}{u_{e}^{s}+\mathcal{K}_{v}/\mathcal{K}_{e}u_{e}^{v}}.
\end{equation}
In this case, the proper distance can be evaluated as
\begin{eqnarray}\label{35}
dl_{p}=\mathcal{K}_{\tau}u_{e}^{\tau}ds =
\frac{\mathcal{K}_{s}}{g|\mathcal{K}_{v}|}dr,
\end{eqnarray}
where $\tau$ represents the photon path along the affine parameter
$s$. Here, we also consider that the emissive specific intensity is
monochromatic, so Eq. (\ref{26}) is still valid and hence, the
infalling spherical accretion can be calculated as
\begin{equation}\label{36}
I^{\text{obs}}_{\star}(b)\propto\int
g^{3}\mathcal{K}_{s}(r^{2}|\mathcal{K}_{v}|)^{-1}dr.
\end{equation}
\begin{figure}[thpb]\centering
\includegraphics[width=5.9cm,height=4.6cm]{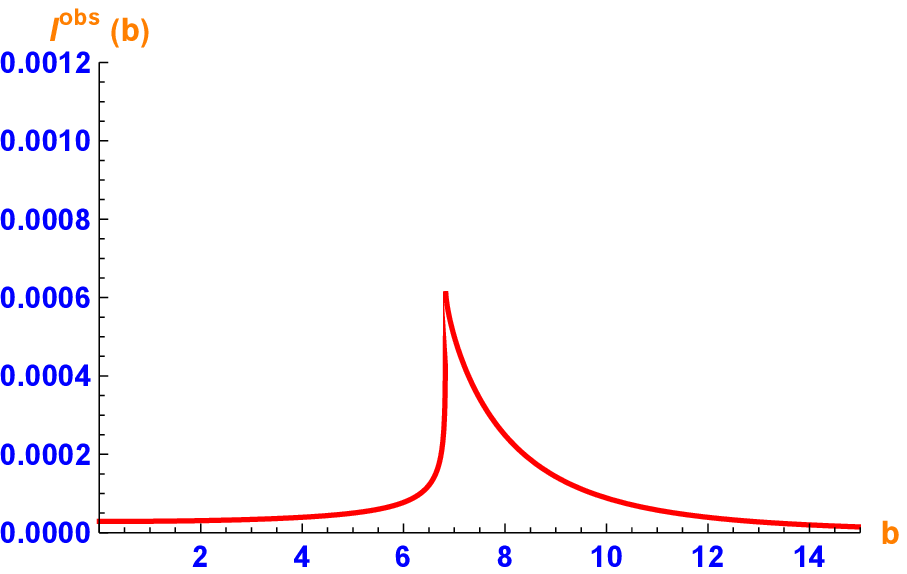}
\includegraphics[width=5.9cm,height=4.6cm]{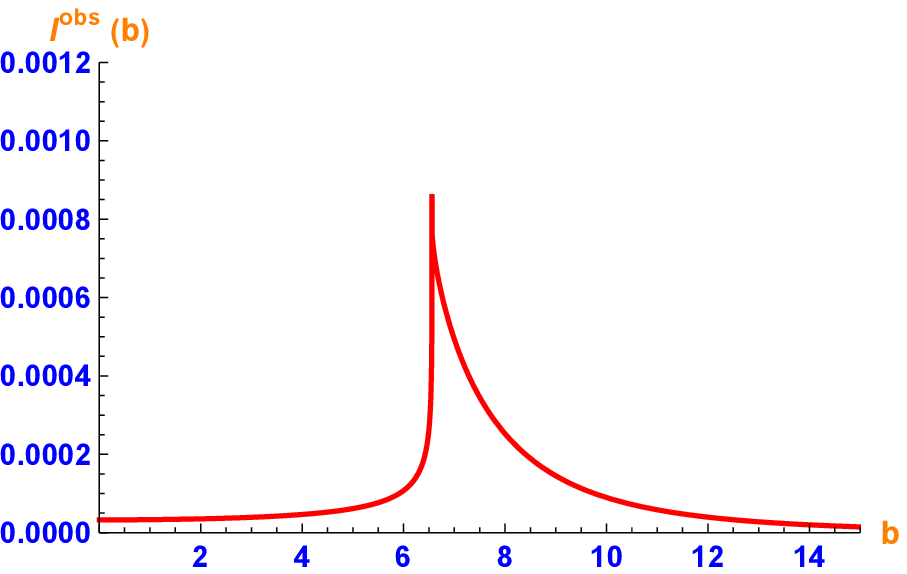}
\includegraphics[width=5.9cm,height=4.6cm]{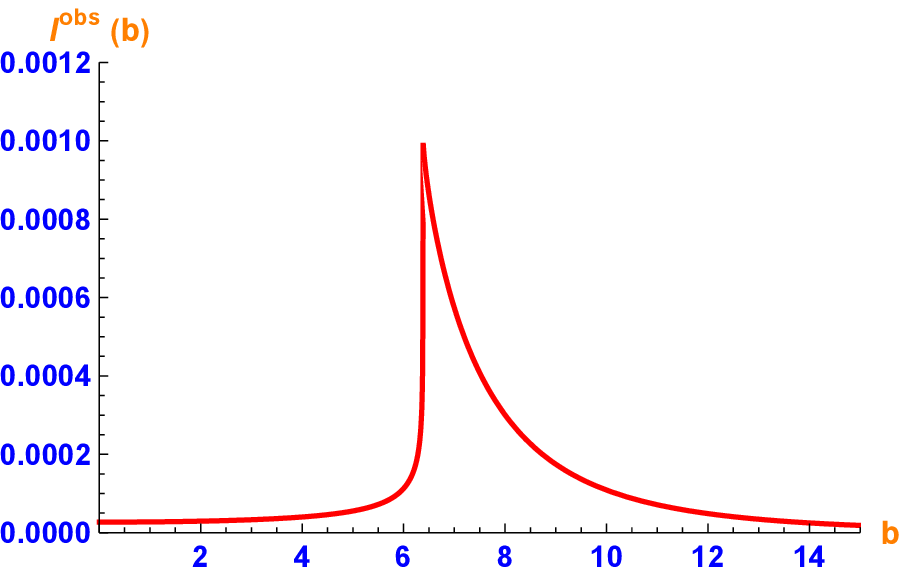}
\includegraphics[width=5.9cm,height=5.5cm]{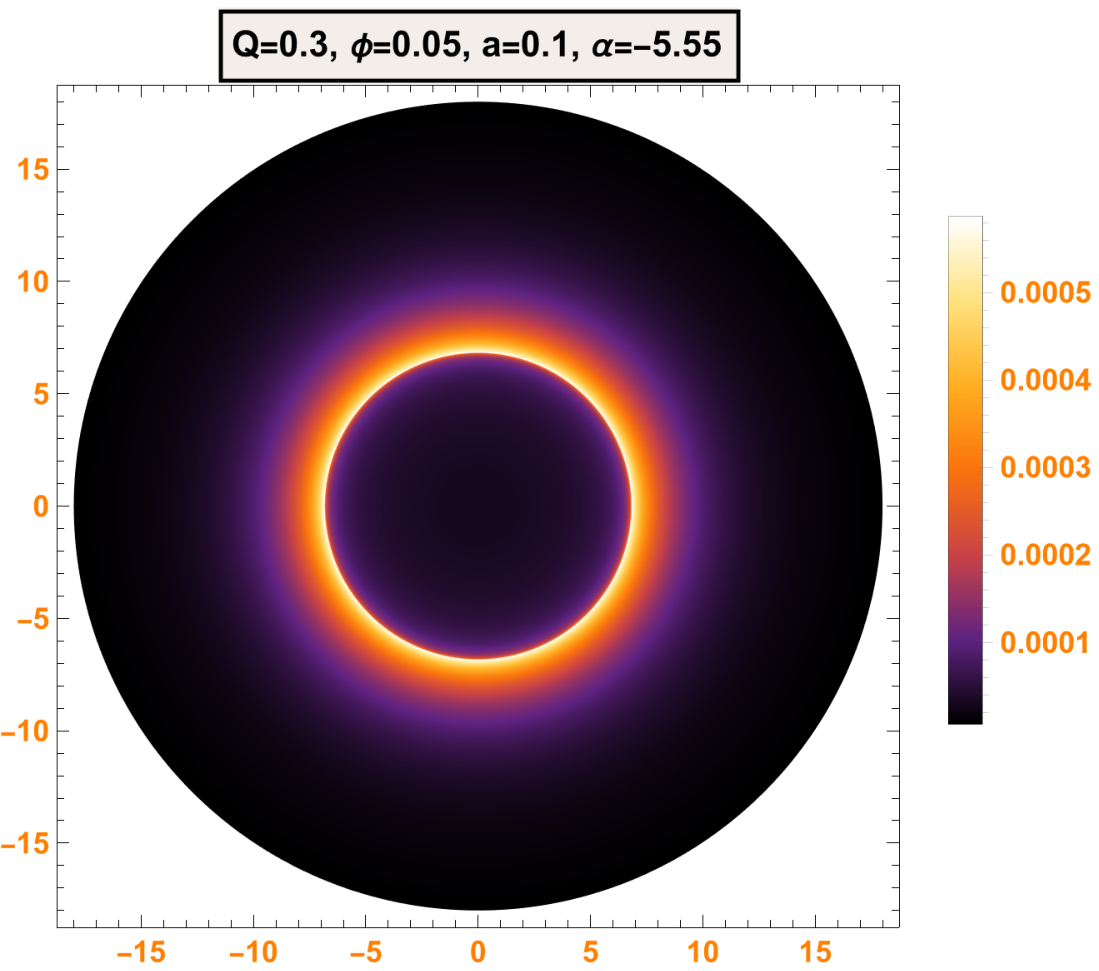}
\includegraphics[width=5.9cm,height=5.5cm]{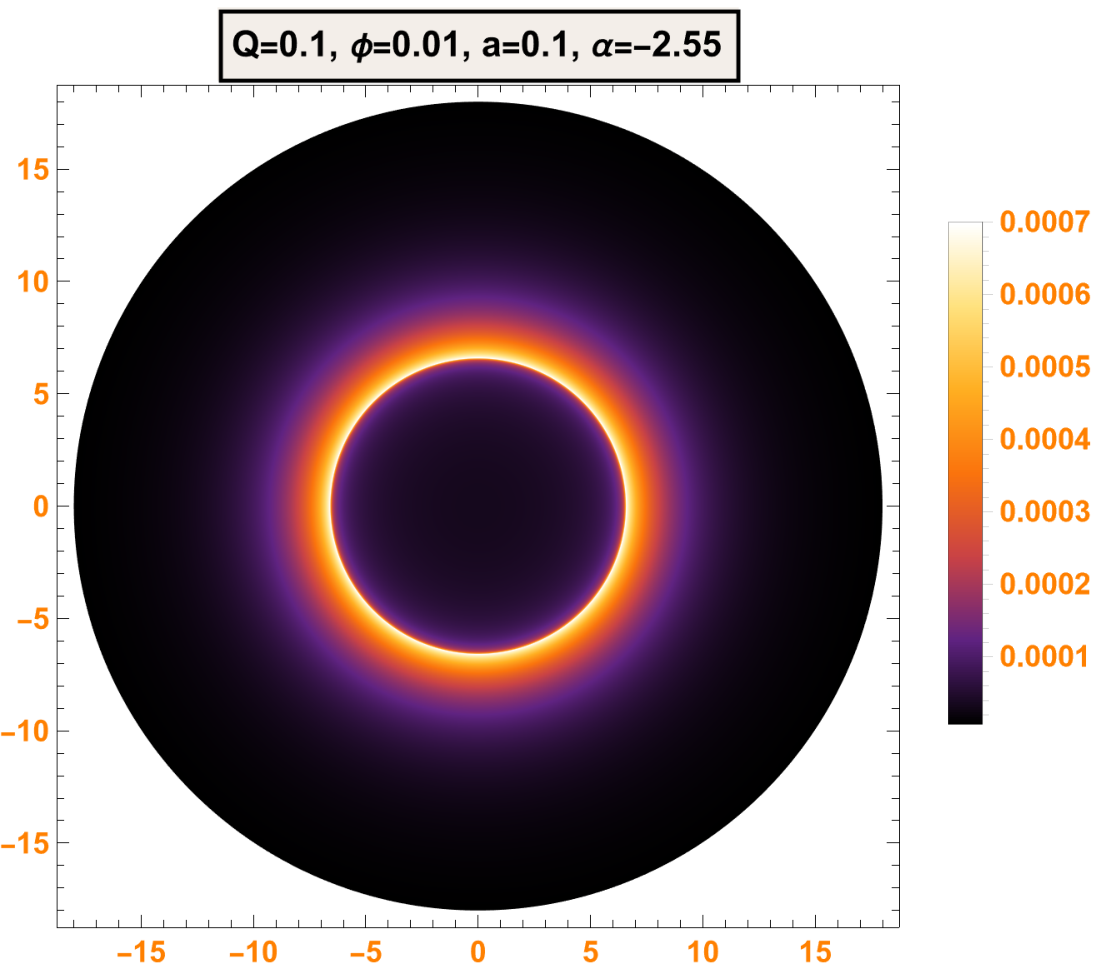}
\includegraphics[width=5.9cm,height=5.5cm]{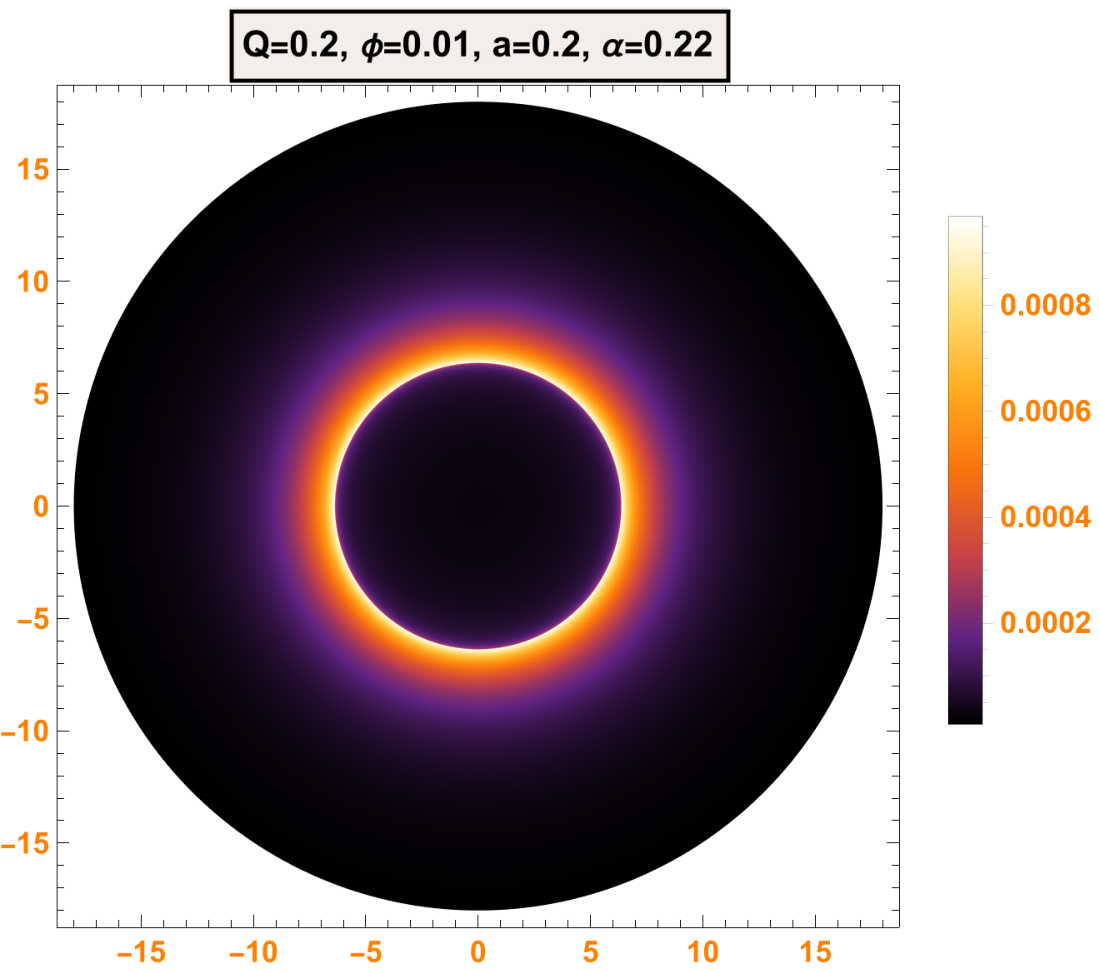}
\caption{The graphical interpretation of the total observed
intensity $I^{\text{obs}}_{\star}(b)$ reflects with an infalling
spherical accretion flow matter, for different BH state parameters
and the mass of BH as $M=1$. We choose these numerical values as
three examples, and one can find a comprehensive details in the main
texts.}
\end{figure}
From Eq. (\ref{36}), we investigate the shadow of the considering BH
by a static observer for different model parameters with infalling
accretion. In Eq. (\ref{36}), an absolute value of $\mathcal{K}_{v}$
is included, which represents that with the change of photon's
direction of motion, the sign of $\mathcal{K}_{v}$ also changed.
From Fig. \textbf{9} (top row), we observed that the intensity of
light increases significantly first as well to peak $b=b_{p}$, and
then booster down to lower values. The observational features are
relatively equivalent to that in the rest accretion case as we
discussed earlier for Fig. \textbf{8}.

The two-dimensional observational appearance of BH shadow image is
shown in Fig. \textbf{9} (bottom row). We noticed that the
resolution of the BH shadow and the location of the photon sphere
are the similar phenomenological prescription as those with the
static accretion flow. However, a major new disks viewed is found
that in the central region, the infalling BH shadow has typically
darker contribution in space-time properties as compared to that
with static accretion, which is presumably accounted due to Doppler
effect which relates the matter orbiting around the jet axis. More
significantly, one can observe that the influence of this
contribution is more prominent, which is close to the event horizon
of the considering BH system.
\section{Conclusions and Discussion}

During the last two decades, the analysis of shadow cast by the astrophysical stellar objects, especially for BH and wormholes have been obtained significant attention in many research areas. The first detection of a BH shadow found at the hearts of giant galaxies from the EHT collaboration led to the more realistic interpretation of these astrophysical objects in nature. In the present study, we mainly investigate the shadows and photon spheres of BH, which is illuminated by a thin accretion disk. In this scenario, we have considered the optical appearance of four-dimensional Gauss-Bonnet BH in the presence of charge along with the cloud of strings and NC geometry of BH distribution.

For the viability of our developed structure in the framework of our considering BH shadow formulation, we have examined some physical properties under the influence of model parameters such as effective potential, deflection of light near the BH, shadows, and photon rings with optically and geometrically thin disk emission and observed specific intensities with static and infalling accretion flow models. The physical significance of the discussed properties for some specific choices of model parameters for our considered systems is listed below.
\begin{itemize}
\item Effective Potential: Using the effect of null geodesic, we
depicted the behavior of effective potential and photon ring orbit in
this geometry. From Fig. \textbf{1}, one can see that the
trajectories of effective potential significantly vary with
the variations of parameters. Particularly, the value of the event
horizon $r_{h}$, critical curve $b_{c}$ and photon sphere $r_{p}$
decreases with the increase of $\alpha$ and these quantities
directly increase with increasing values of $a$. In addition, we also depicted the ingoing and
outgoing trajectories of light rays corresponding to effective
potential as shown in Figs. \textbf{2} and \textbf{3}. For
instant, we depicted here the behavior of effective potential just as
example for some specific choices of model parameters and
investigated the dynamics of the photon sphere under the influence of
these values.

\item Light Bending Near a BH: The optical appearance of accretion matter near the
BH from a static observer's eye can be understood by tracing the
null geodesics. To understand the total change in the optical plane,
we take three optical and geometrical accretion flow models as
examples of null geodesic light trajectories with impact parameter
$b$. We plotted this accretion in two different ways in Fig.
\textbf{4}. The top row of Fig. \textbf{4} shows the light ray
trajectories according to the total number of orbits,
$n(b)=\varphi/2\pi$. The motion of light rays in a straight line
would correspond to $n(b)=1/2$, and the bending of light rays that
do not enter the BH is calculated by $n(b)-1/2$. The singularity of
these plots lie at $b=6.82781M$ (left panel), $b=6.55952M$ (Middle
panel) and $b=6.38135M$ (right panel).

The bottom row of Fig. \textbf{4} gives a clearer picture
of what a static observer would see at large distances. The light
rays can be classified into three regions such as direct emission,
lensing ring and photon ring according to the number of distinctions
at the equatorial plane. Despite these distinct configurations in the
total luminosity of BH solutions, there is not only a dark interior
the region, but also the direct emission makes a major contribution to the
brightness while the lensing ring makes a minor contribution and the
role of photon ring in observational appearance can be safely
ignored regarding their negligible contribution.

\item Transfer Functions: We try to understand the emission of
light rays originating near the BH from an optically thin disk with
the formulation of specific intensity depending on the radial
coordinate. We plotted the \textit{first three transfer functions}
in Fig. \textbf{5}. There is no light deflection inside the radius
because none of the transfer functions has support for $b\leq2.9M$.
The slope of the \textit{first transfer function} lies in the entire
range, so the direct image represents the red-shift source file.

The \textit{second transfer function} supports the lensing ring and
the image profile here is highly (de)magnified of the back side of
the disk, while the \textit{third transfer function} represents the
photon ring, where the image is extremely demagnetized on the front
side of the disk. This image profile as well as further images are
highly demagnified, so, the contribution of these images is
negligible to the total luminosity.
\item Observational Features of BH: In addition, we considered three
toy models of optically and geometrically thin disk emission flow to
further study the observational appearance and then compared the
observed specific intensities. The emitted intensity observed from
these three models peaks at the isco for time-like observers, and at
the unstable circular orbit for photons spheres and close the
horizon for BH solution. The simulation obtained from specific
intensities with the help of these three models yield different
qualitative emissions and physical scenarios.

From Figs. \textbf{6} and \textbf{7}, the photon ring is an
extremely curved light ray and has so narrow area and hence, the
participation to the total flux can be negligible. The region of the
lensing ring is wider than the photon ring and made a large
contribution to the total flux. But this contribution is very small
as compared to direct emission flow case. Hence, our obtained
solutions showed that the role of direct emission is always
appreciable to the total observed intensity.

\item Specific Intensity with Static and Infalling Spherical
Accretion: From the observed intensity, one can see that there is a
bright sphere ring outside the central region and this brightness
gradually varies with the influence of BH state parameters. In the
case of infalling accretion, we found that the interior region of
the shadow has lower observation luminosity than that of the static
case, as one can see from the density map as depicted in Figs.
\textbf{8} and \textbf{9}. Furthermore, the static model has
significantly lower trajectories than the infalling accretion one.
So, the Doppler effect of infalling matter caused the most prominent
contrast between these two accreting models. However, our
considering BH parameters would change the position/location of the
photon sphere BH profiles.
\end{itemize}

We obtained these results for a more realistic model as compared to \cite{14,15} and tried to
enhance the brightness of the photon ring and BH shadow. We expect that these results inspire
the theoretical study of BH shadow image and other physical quantities that may be
fruitful for the observational teams working on achieving high resolution of the photon sphere
and BH accreting matter configuration in GR as well as other modified theories of gravity.

\end{document}